\documentclass[12pt]{article}
\usepackage{aaspp4}
\usepackage{flushrt}
\usepackage{graphicx}
\usepackage{psfrag}
\tighten
\def\ch#1{{#1}}
\def\eV{{\;{\rm eV}}}
\def\GZK{{\rm GZK}}
\def\kms{{\rm kms}${}^{-1}$}

\def\msun{\,{\rm M}_\odot}

\def\rc{r_{\rm c}}
\def\repsilon{r_\epsilon}
\def\erf{{\rm erf}}
\def\rs{r_{\rm s}}
\def\rc{r_{\rm c}}
\def\Rp{{R^\prime}}
\def\zp{{z^\prime}}
\def\xp{{x^\prime}}
\def\yp{{y^\prime}}
\def\bs{{\bf s}}
\def\br{{\bf r}}
\def\rsun{{\bf R_\odot}}
\def\rmax{{r_{\rm max}}}
\def\smax{{s_{\rm max}}}

\def\atan{{\rm atan}}
\def\acos{{\rm acos}}
\def\R{{\cal R}}
\def\S{{\cal S}}
\def\ffrac#1#2{{\textstyle\frac{#1}{#2}}}
\begin{document}
\title{The Anisotropy of the Ultra-High Energy Cosmic Rays}
\author{N.W. Evans~$^1$, F. Ferrer~$^2$ and S. Sarkar~$^1$}
\medskip

\affil{$^1$Theoretical Physics, University of Oxford,  1 Keble
Road, \\ Oxford, OX1 3NP, ENGLAND}

\medskip

\affil{$^2$Grup de F{\'\i}sica Te{\`o}rica and Institut de F{\'\i}sica d'Altes
Energies, Universitat Aut{\`o}noma de Barcelona, 08193 Bellaterra,
SPAIN}

\begin{abstract}
Ultra-high energy cosmic rays (UHECRs) may originate from the decay of
massive relic particles in the dark halo of the Galaxy, or they may be
produced in nearby galaxies, for example by supermassive black holes
in their nuclei.  The anisotropy in the arrival directions is studied
in four dark halo models (cusped, isothermal, triaxial and tilted) and
in four galaxy samples (galaxies intrinsically brighter than
Centaurus~A within $50$ and $100$ Mpc, and galaxies intrinsically
brighter than M32 within $50$ and $100$ Mpc).

In decaying dark matter models, the amplitude of the anisotropy is
controlled by the size of the Galactic halo, while the phase is
controlled by the shape. As seen in the northern hemisphere, the
amplitude is $\sim 0.5$ for cusped haloes, but falls to $\sim 0.3$ for
isothermal haloes with realistic core radii. The phase points in the
direction of the Galactic Centre, with deviations of $\sim 30^\circ$
possible for triaxial and tilted haloes.  The effect of the halo of
M31 is too weak to provide conclusive evidence for the decaying dark
matter origin of UHECRs.  In extragalactic models, samples of galaxies
brighter than Centaurus~A produce substantial anisotropies ($\sim
1.8$), much larger than the limits set by the available data.  If all
galaxies brighter than M32 contribute, then the anisotropy is more
modest ($\lesssim 0.5$) and is directed toward mass concentrations in
the supergalactic plane, like the Virgo cluster.

Predictions are made for the south station (Malarg{\"u}e) of the Pierre
Auger Observatory.  If the UHECRs have a Galactic origin, then the
phase points towards the Galactic Centre.  If they have an
extragalactic origin, then it points in the rough direction of the
Fornax cluster. This provides a robust discriminant between the
two theories and requires $\sim 350-500$ events at South Auger.
\end{abstract}

\bigskip
\noindent 
{PACS classification codes: 96.40, 98.70.S, 95.35, 98.35}

\section{Introduction}

Cosmic ray particles with energy in excess of $\sim 4 \times
10^{19}\eV$ have been detected by a number of independent experiments
over the last few decades [\cite{review}]. These include the Haverah
Park [\cite{hp}], Fly's Eye [\cite{fe1}] and Akeno Giant Air Shower
Array (AGASA) [\cite{agasa1}] experiments, which together have
recorded over a hundred such ultra-high energy cosmic rays
(UHECRs). It has long been recognised that the existence of these
UHECRs poses an awkward problem [\cite{gzk}]. At such high energies,
the typical range of protons decreases rapidly because of interactions
with the cosmic microwave background photons, becoming as low as $\sim
20$~Mpc at the highest observed energy of $\sim 3 \times 10^{20}$ eV
[\cite{sempr}]. It is even smaller for nuclei [\cite{er}].  There is
evidence for a change in composition towards protons at the highest
energies from the elongation rate of the showers observed by Fly's Eye
[\cite{fe1}] and its successor HiRES [\cite{hires}] as well as AGASA
(when appropriately reanalysed) [\cite{dawson}].  If the UHECRs are
nucleons and their sources are distributed homogeneously throughout
the universe, this should create a ``GZK cutoff'' in the energy
spectrum at $E_\GZK\sim 4\times10^{19}\eV$ [\cite{sempr}]. This can be
evaded if the UHECRs are neutral particles, e.g. neutrinos or
photons. However, neutrinos which interact only weakly have too small
a cross-section to initiate the observed airshowers
[\cite{nu}].\footnote{If new physics, such as large extra dimensions
(implying TeV scale quantum gravity), is invoked, the neutrinos may
interact more strongly through Kaluza-Klein graviton exchange. It
appears unlikely that the increased cross-section would be adequate to
create the observed air-showers [\cite{foot}]} Photon-initiated
showers would have different characteristics than those observed; in
particular, the highest energy Fly's Eye event was probably a
nucleon [\cite{fe1,hvsv95}]. A recent analysis of horizontal showers
in the Haverah Park data sets a bound of 55\% on the photon component
(and a bound of 30 \% on the iron nuclei content) of post-GZK UHECRs
[\cite{ahvwz00}]. So, if cosmic rays with energies $E>E_\GZK$ are
protons, as we assume on the basis of the above evidence
[\cite{fe1,dawson,hvsv95}], then they they must originate within the
Local Supercluster.

There has been a plethora of suggestions as to the possible
origin of UHECRs, which may be divided into `top-down' and `bottom-up'
models. In the former class, the extreme energies are provided by the
decay of relic topological defects or super-massive particles.  In the
latter class, the extreme energies are provided by the acceleration of
particles in astrophysiscal sites such as gamma-ray bursts and active
galaxies. In this paper, we investigate the observational signatures
of one `top-down' and one `bottom-up' model in some detail.

One `top-down' possibility is that the UHECRs may originate from the
decay of super-massive relic particles which may constitute (a
fraction of) the dark matter in galactic haloes
[\cite{bkv97,bs98}]. Such particles must have a mass
$\gtrsim10^{11}$~GeV to account for the highest energy events, with a
lifetime exceeding the age of the universe. A well-motivated particle
physics candidate with both the required mass and metastability,
proposed {\em before} the definitive Fly's Eye event [\cite{fe1}], is
the ``crypton'' --- the analogue of a hadron in the hidden sector of
supersymmetry breaking in string theories [\cite{crypton}]. It has
been recently noted that such particles can be readily produced with a
cosmologically interesting abundance in the time-varying gravitational
field at the end of inflation [\cite{gravprod}].  A key test of this
model is the expected small anisotropy caused by the offset of the Sun
from the centre of the Galaxy [\cite{dt98}].  It has been claimed that
the expected anisotropy is already higher than that observed at lower
energies $\sim(1-5)\times10^{18}\eV$ [\cite{bsw99}]. However, this is
beside the point, as such measurements do not constrain the new `flat'
spectrum component of cosmic rays which extends beyond the GZK cutoff
[\cite{mtw99}]. It has also been claimed that the absence of any
excess signature in the direction of the halo of the nearby Andromeda
galaxy is an ``insuperable'' problem for this model
[\cite{bsw99}]. This claim has already been contested on the grounds
that any excess signal would not be detectable in the present sample
of $\sim 0.5$ events per $10^\circ \times 10^\circ$ solid angle
[\cite{mtw99}].  A number of authors [\cite{dt98}-\cite{bm98}] have
calculated the expected anisotropy assuming either an isothermal halo
profile or the ``universal'' density profile suggested for cold dark
matter dominated halos [\cite{nfw96}].  However, very little is known
for certain about the shape, profile and extent of the dark halo of
the Galaxy, and so it is important to examine more general halo
models.

One `bottom-up' possibility is that the UHECRs originate in nearby
extragalactic sources. Estimates of the intergalactic magnetic field
(IGMF) are typically of ${\cal O}(10^{-9})$~Gauss [\cite{igmf}], which
is insufficient to deflect such energetic protons by more than a few
degrees. Hence, the UHECRs should point back in the direction of the
sources. Nevertheless, searches for sources within $\sim3^\circ$ of
observed events have not had any success [\cite{es95}]. The UHECRs
(particularly in the Haverah Park sample [\cite{hp}]) show a
correlation with the supergalactic plane [\cite{supergal}], but this
is not seen in the Fly's Eye [\cite{fe2}] or AGASA [\cite{agasa2}]
data. The latter dataset has the largest number of events and a
detailed analysis finds the arrival directions are consistent with
isotropy, but with some evidence of clustering on an angular scale of
$\sim2.5^\circ$ close to the supergalactic plane
[\cite{agasa2}]. Various possibilities have been suggested for
altering the IGMF to enable an origin in nearby radio galaxies or
active galactic nuclei, e.g., M87 or Centaurus~A [\cite{rubbish}], but
these have difficulties [\cite{d00}].  For UHECRs to originate in
nearby extragalactic sources, they must come from a population that is
larger than the nearest active galactic nuclei or radio galaxies.  It
is believed that all big galaxies harbour supermassive ($\sim 10^{6} -
10^{9} \msun$) black holes and have passed through an active phase in
the past, even if they are quiescent today [\cite{rees}]. Processes
are known that permit efficient energy extraction from supermassive
black holes [\cite{bz}], and so one possibility is that nearby
galactic nuclei are the arena of production of UHECRs [\cite{extra}],
even though detailed pathways are lacking.  Studies have already shown
that if the sources of UHECRs trace the density field of nearby
galaxies, then the observed flux can possibly be recovered, at least
for hard injection spectra [\cite{olinto}]. Although the energetics of
the production of the particles may be problematic, this is
nonetheless an interesting model to contrast with the decaying dark
matter hypothesis. It also gives rise to an anisotropy signal in the
arrival directions of UHECRs, as the nearby rich clusters (like Virgo
and Fornax) make a disproportionately large contribution. We must keep
in mind however that the expected anisotropy may be diluted by a
factor of up to $\sim 2$ by the concomitant isotropic background from
even further sources. In practice, given that this dilution is
sensitive to the assumed injection spectrum of the sources (e.g., it
would be negligible for an $E^{-3}$ spectrum [\cite{extramt}]), we do
not explicitly take this into account.

In this paper, we always neglect the effects of magnetic fields on
UHECR trajectories. This is justifiable if the sources lie in the
Galaxy's halo, as studies suggest that the influence of the Galactic
magnetic field on trajectories is small if the energies are greater
than $4 \times 10^{19}$ eV [\cite{ptuskin}].  It is a more
controversial assumption if the sources are extragalactic.  If an
inhomogenous energy density distribution which follows the
Lyman-$\alpha$ forest distribution is assumed [\cite{bbo}], then the
estimates of the IGMF can be as large as ${\cal O}(10^{-8})$~Gauss.
It also possible that the IGMF has a small filling factor and is
structured on large scales of up to tens of Mpcs [\cite{rkb}]. Hence,
it is sometimes claimed that a plausible upper limit to the IGMF could
be as high as ${\cal O}(10^{-6})$~Gauss, in which case considerable
deflection of incoming charged UHECRs is a real possibility
[\cite{mt}]. However, there are also strong arguments against such
high values for the IGMF [\cite{wax}]. In particular, it is hard to
see how such a field can be generated dynamically over tens of Mpcs,
as the eddy turnaround time is larger than a Hubble time. Note that
our choice of $4 \times 10^{19}$ eV as the threshold is a reasonable
compromise, i.e., it is high enough to minimize magnetic deflection
without losing too much flux.  In principle, a lower energy cut may
appear attractive for analysing the halo source model, since the
propagation distance is less than $\sim 100$ kpc. However, to fit the
spectrum still requires the new `flat' component of cosmic rays (from
dark matter decays in this case) to remain subdominant below the
$E^{-2.8}$ power-law extension from lower energies until $\sim4\times
10^{19}$ eV [\cite{bs98,st01}]. Moreover, this cut enables easy
comparison with earlier studies of the expected anisotropy
[\cite{hayashida,uchihori}].

The motivation for the paper is that the prospects for detection of
any anisotropy signal in the UHECRs in the next few years are good.
The southern station of the Pierre Auger Observatory [\cite{auger}] is
already under construction at Malarg{\"u}e, Argentina and will be
complete by 2004. The array consists of 1600 water-Cerenkov detectors
distributed in a grid covering about 3000 square kilometers. The
northern station is planned for Utah and will then enable continuous,
full-sky coverage. Estimates have already been made of the number of
events that Auger will detect after five years of operation
[\cite{ps}]. It is reckoned that there will be $\sim 2200$ above $4
\times 10^{19}$ eV, $\sim 250$ above $10^{20}$ eV and $\sim 35$ above
the highest energy Fly's Eye event ($3.2 \times 10^{20}$
eV). Another experiment that will have a considerable impact is
the Extreme Universe Space Observatory~[\cite{euso}]. This is
scheduled for flight on the International Space Station starting in
mid 2007.  It will detect flourescent light produced when UHECRs
interact with the Earth's atmosphere.  Given the richness of
forthcoming experimental opportunities, it is interesting to examine
possible tests. Can we use the anisotropy signal to distinguish
between a local origin in the halo of the Galaxy and a more distant
origin in nearby galaxies?  If the UHECRs originate in the
Galaxy's halo, what can be learnt about its structure? If the UHECRs
emanate from nearby astrophysical sources, what can be learnt about
their distribution?

The paper is organised as follows. Section 2 discusses the theoretical
framework for analysis of the anisotropy signal.  Section 3 introduces
four halo models, while Section 4 constructs four samples of nearby
galaxies. The prospects for distinguishing between the galactic and
extragalactic origins of UHECRs with the existing Haverah Park and
AGASA experiments and especially the southern station of the Auger
Observatory are examined. Section 5 sums up our conclusions.
 

\section{The Anisotropy Signal}

\subsection{The Detector Response}

The emissivity of UHECR per unit volume is proportional to the number
density of sources $n(\br)$. The incoming flux of UHECR per unit solid
angle $F$ as a function of right ascension $\alpha$ and declination
$\delta$ is
\begin{equation}
F(\alpha, \delta) \propto \int_0^{\smax (\alpha, \delta)}
ds\,n(\rsun + \bs(\alpha, \delta)),
\end{equation}
where $\bs$ is the heliocentric position vector. Here, $\smax$ marks
the extent of the distribution of sources, which may depend on
direction.

The measured flux of UHECR is the incoming flux modulated by the
response of the detector $h(\delta)$, which is the relative efficiency
of the detection of events with direction. It depends only on the
declination and not on the right ascension, because, if the detector
is run with reasonable efficiency, there will be almost uniform
exposure in right ascension after a year or so.  The response as a
function of declination depends partly on the attenuation of showers
in the atmosphere, partly on the sensitivity of the detector as a
function of zenith angle.

For the scintillator arrays of AGASA and the water-Cerenkov detectors
used at Haverah Park, the declination distribution $h(\delta)$ of the
observed events is reproduced in Figure 1 of Uchihori et
al. [\cite{uchihori}].  Our procedure is to fit third-degree
polynomials (conveniently bounded by step functions) to these
experimental curves. For AGASA, this gives
\begin{equation}
h_{\rm AG}(\delta) = 0.323616 + 0.0361515\,\delta - 5.04019 \times 10^{-4}
\,\delta^2 + 5.53941 \times 10^{-7}\,\delta^3,
\end{equation}
valid for $ -8.0^\circ < \delta < 87.5^\circ$. For Haverah
Park, this gives
\begin{equation}
h_{\rm HP}(\delta) = 0.108234 + 0.0320955\,\delta - 1.881358 \times
10^{-4}\,\delta^2 -1.948817 \times 10^{-6}\,\delta^3,
\end{equation}
valid for $\delta > -3.3^\circ$. These expressions are normalised to a
maximum value of unity.

In the absence of measurements from South Auger, we explore two
possibilities. First, following Medina Tanco \& Watson [\cite{mtw99}],
we use the declination distribution found at Haverah Park ($54^\circ$
N), mirrored and shifted to the latitude of the South Auger site at
Malarg{\"u}e, Argentina ($35^\circ$ S). The motivation for this is
that the peak of the response function usually coincides with the
geographic latitude (at least when there is not continuous exposure at
the Poles). So, we use:
\begin{equation}
h_1(\delta) = h_{\rm HP}(19^\circ - \delta).
\end{equation}
Second, following Sommers [\cite{ps}], we use the analytic expression
valid for a detector at a single site with continuous operation and
constant exposure in right ascension.  Suppose the detector is at
latitude $\lambda$ and that it is fully efficient for particles
arriving with zenith angles $\theta$ less than some maximum value
$\theta_{\rm m}$, which is taken as $60^\circ$. This gives us:
\begin{equation}
h_2(\delta) \propto\ \cos\lambda \cos\delta \sin\alpha_{\rm m} +
        \alpha_{\rm m} \sin\lambda \sin\delta,
\end{equation}
where $\alpha_{\rm m}$ is the local hour angle at which the zenith angle 
becomes equal to $\theta_{\rm m}$. It is given by
\begin{eqnarray} 
\alpha_m = \left\{ \begin{array}{ll} 
\ 0 & \mbox{if $\xi > 1$}\\
\ \pi & \mbox{if $\xi < -1$}\\
\acos (\xi) & \mbox{otherwise}
\end{array} \right. 
\end{eqnarray}
and
\begin{equation}
\xi \equiv
\frac{\cos \theta_{\rm m} -
\sin \lambda \sin \delta}{\cos \lambda \cos \delta}.
\end{equation}
Note that the two declination distributions $h_1(\delta)$ and
$h_2(\delta)$ are quite different. The maximum of $h_2(\delta)$ is
(generally) in the direction of the Pole, whereas the maximum of
$h_1(\delta)$ is in the direction of the geographic latitude of the
South Auger site.  Often, but not always, our results do not depend
too much on the details of the response function.

\begin{figure}
\begin{center}
\begin{tabular}{c@{\hspace{2cm}}c}
  \psfrag{r}{$\xi$}
  \psfrag{p}{$P(\xi)$}
  \includegraphics[width=6.5cm,height=6cm]{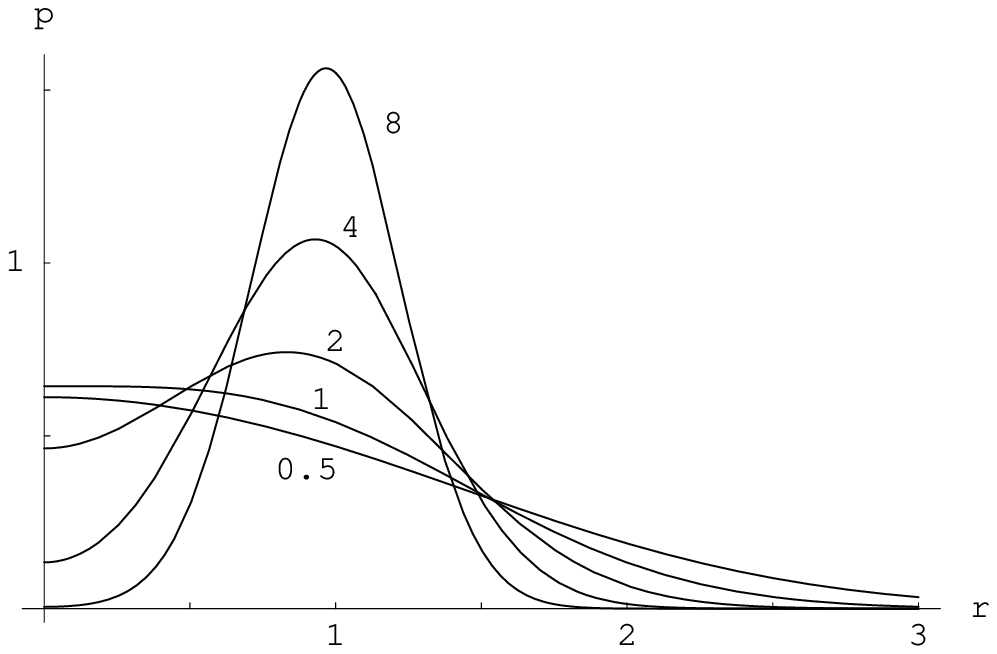}&
  \psfrag{r}{$\vartheta$}
  \psfrag{p}{$P(\vartheta)$}
  \psfrag{m}{$\scriptstyle -\pi$}
  \psfrag{n}{$\scriptstyle -\pi/2$}
  \psfrag{s}{$\scriptstyle \pi$}
  \psfrag{q}{$\scriptstyle \pi/2$}
  \includegraphics[width=6.5cm,height=6cm]{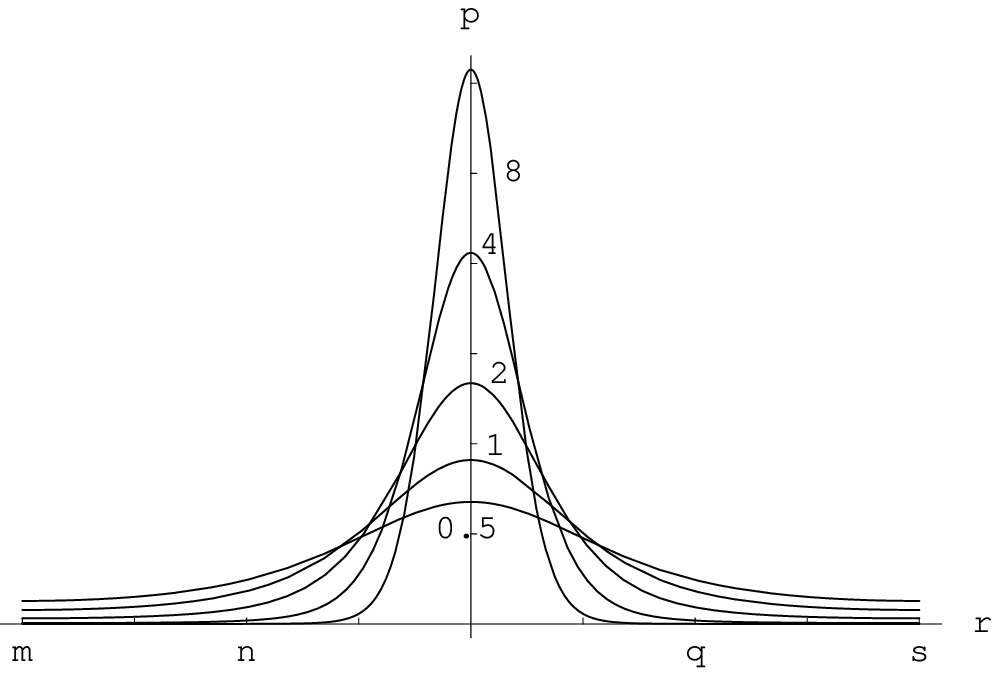}\\
(a) & (b) 
\end{tabular}
\end{center}
\caption{Plots of the prob bility distributions $P(\xi)$, where
$\xi=\S/\R$, (left panel) and $P(\vartheta)$ (right panel) as given by
eqns (\ref{eq:distsa}) and (\ref{eq:distsb}) respectively. The curves
are labelled with the relevant value of $k_0 = \ffrac{1}{4}\R^2
N$. Directly measurable quantities are only involved through this
quantity.}
\label{fig:distributions}
\end{figure}
\begin{table}
\begin{center}
\begin{tabular}{|l|c|c|c|c|c|c|c|} \hline
Experiment & Amplitude & Phase & $N$ & $\sigma_\S^+$ & $\sigma_\S^-$ 
& $\sigma_\S$ & $\sigma_\vartheta$ \\
& $\R$ & $\psi$ & \null & ($95 \%$) & ($95 \%$) &
($95 \%$) & ($95 \%$) \\ \hline 
AGASA & $0.262$ & $298.5^\circ$ & $57$  & $\infty$ & 
$0.209$ & $0.242$ & $145^\circ$\\ \hline 
Haverah Park& $0.583$ & $173.6^\circ$ & $27$ & $0.622$ 
& $0.452$ &$0.491$ & $95^\circ$ \\ \hline \hline
Auger I& $0.262$ & $298.5^\circ$ & $1000$ & $0.09$ &
$0.09$ & $0.09$ & $19.6^\circ$ \\ \hline
Auger II& $0.583$ & $173.6^\circ$ & $1000$ & $0.09$ &
$0.09$ & $0.09$ & $6.5^\circ$ \\ \hline
\end{tabular}
\end{center}
\caption{Summary of the experimental data for AGASA and Haverah Park,
together with two projections for Auger. The first projection assumes
that the \ch{amplitude and phase are} the same as that seen by AGASA,
the second the same as that seen by Haverah Park; in both cases the
number of events is increased to 1000. The amplitude $\R$ and phase
$\psi$ are calculated from the number $N$ of UHECRs, together with the
$95 \%$ confidence limits. (For AGASA, the $95 \%$ confidence limit
$\sigma_\S^+$ is not strictly speaking well-defined, as the
integration from $\R$ to $\infty$ gives a probability of only
$0.449$)}.
\label{table:tab1}
\end{table}
\begin{table}
\begin{center}
\begin{tabular}{|l|c|c|c|} \hline
Experiment & $P(>\R)$ & $P(0\!<\!\S\!<\!\sigma_\S)$ & 
$P(0\!<\!\S\!<\!\sigma_\S)$ \\
\null & \null & $(68 \%)$ & $(95 \%)$ \\ \hline
AGASA         & $0.377$ & $0.403$ & $0.578$ \\ \hline 
Haverah Park  & $0.101$ & $0.242$ & $0.494$ \\ \hline \hline
Auger I \& II & $<10^{-3}$ & $< 10^{-3}$ & $<10^{-3}$ \\ \hline
\end{tabular}
\end{center}
\caption{Tests for isotropy on the AGASA and Haverah Park datasets, as
well as the projected Auger datasets. The first column gives the
probability that an isotropic distribution could give an amplitude
greater than that observed $\R$. The second two columns give the
probability that the data comes from a model with theoretical
amplitude $\S$ between isotropy and the $68 \%$ or $95 \% $ symmetric
confidence interval. For Auger I \& II, the probabilities are $<
10^{-3}$, enabling isotropy to be securely ruled out.}
\label{table:tests}
\end{table}
%
\begin{figure}
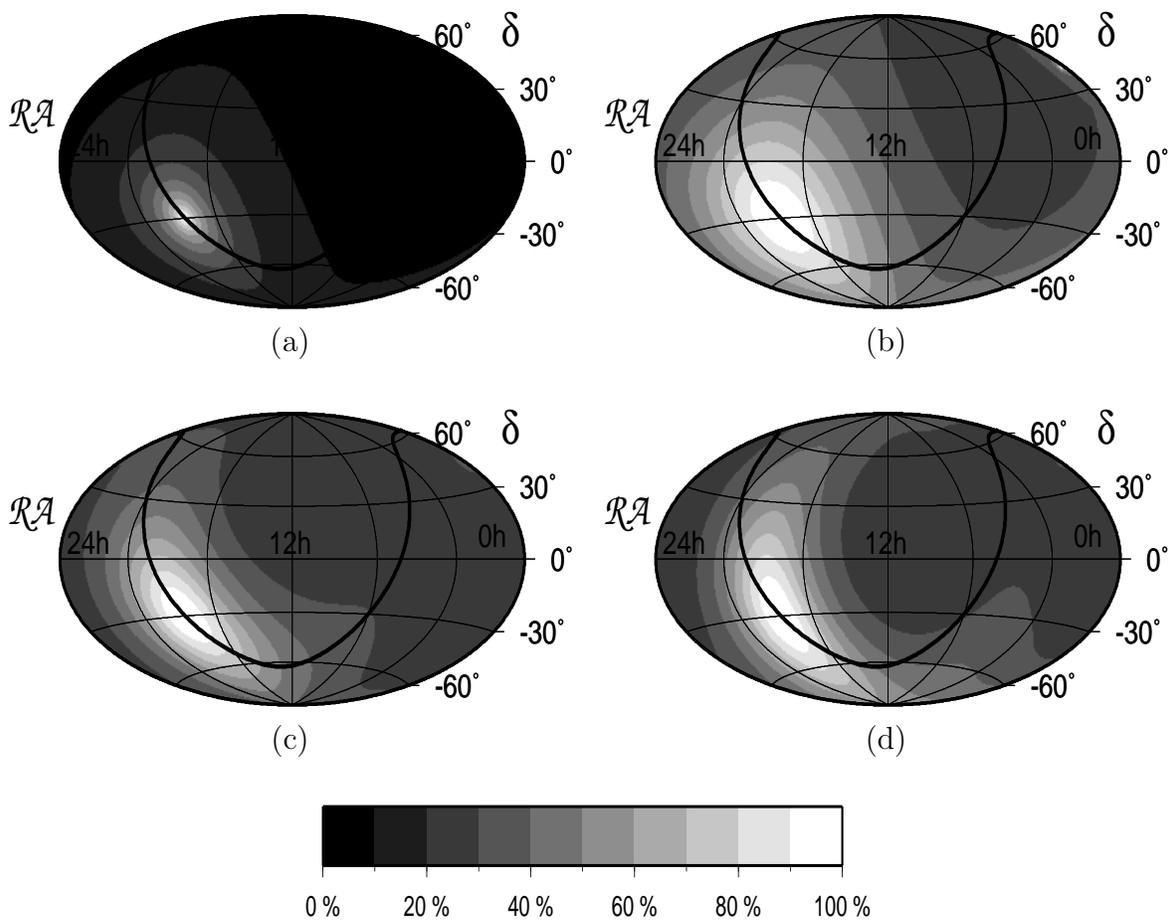

\begin{center}
\begin{tabular}{c}
\begin{tabular}{cc}
\includegraphics[width=7.5cm,height=4cm]{./final_fig2a.eps}&
\includegraphics[width=7.5cm,height=4cm]{./final_fig2b.eps}\\
(a) & (b) \\ &\\
\includegraphics[width=7.5cm,height=4cm]{./final_fig2c.eps}&
\includegraphics[width=7.5cm,height=4cm]{./final_fig2d.eps}\\
(c) & (d)
\end{tabular}\\ \\
\includegraphics[width=7.5cm,height=1.5cm]{./fig2_scale.eps}
\end{tabular}
\end{center}
\caption{Contour plots of the UHECR flux in equatorial coordinates for
our four dark halo models, namely (a) cusped, (b) isothermal, (c)
triaxial and (d) tilted. These are Hamer-Aitoff projections. The
Galactic plane is marked in each figure. The effect of the
halo of M31 is visible in the upper right of each plot.}
\label{fig:allsky}
\end{figure}
\begin{figure}
\begin{center}
\begin{tabular}{c}
\begin{tabular}{cc}
\includegraphics[width=7.5cm,height=4cm]{./final_fig3a.eps}&
\includegraphics[width=7.5cm,height=4cm]{./final_fig3b.eps}\\
(a) & (b) \\ &\\
\includegraphics[width=7.5cm,height=4cm]{./final_fig3c.eps}&
\includegraphics[width=7.5cm,height=4cm]{./final_fig3d.eps}\\
(c) & (d)
\end{tabular}\\ \\
\includegraphics[width=7.5cm,height=1.5cm]{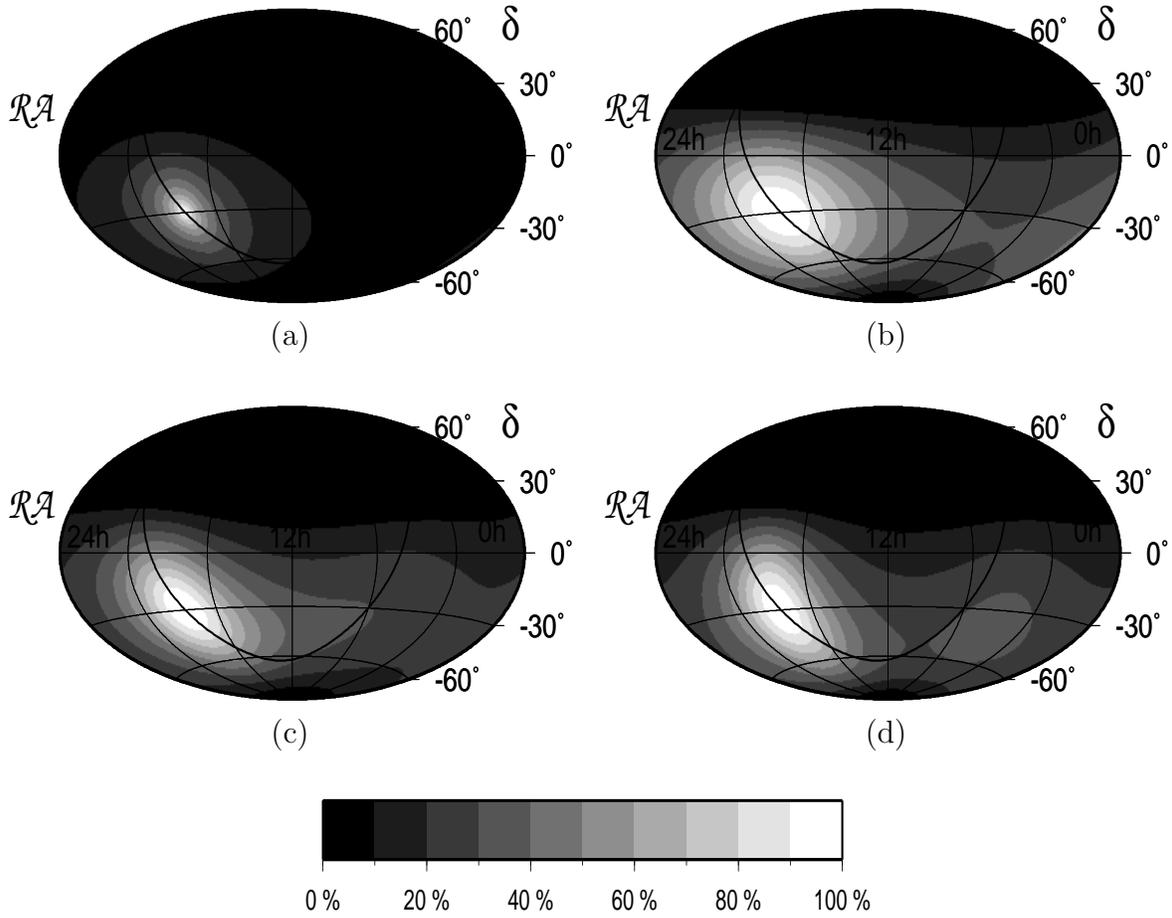}
\end{tabular}
\end{center}
\caption{As Figure~\ref{fig:allsky}, but now the UHECR flux has been
convolved with the response in declination for the southern site of
the Auger Observatory.}
\label{fig:allskyefficiency}
\end{figure}

\subsection{Harmonic Analysis}

The strength and direction of the anisotropy in the arrival directions
of the UHECRs can be quantified by harmonic analysis, as originally
suggested by Linsley [\cite{l75}].  The number of events is
proportional to $N$, where:
\begin{equation}
\label{eq:han0}
N = \int d\delta \int d\alpha \, \cos \delta \, h(\delta) F(\alpha,
\delta).
\end{equation}
Let us define
\begin{eqnarray}
\label{eq:han}
a &=& {2 \over N} 
\int d\delta \int d\alpha \, \cos \delta \cos \alpha \,h(\delta)
F(\alpha, \delta), \nonumber \\
b &=& {2 \over N} 
\int d\delta \int d\alpha \, \cos \delta \sin \alpha \,h(\delta)
F(\alpha, \delta),
\end{eqnarray}
so that the amplitude $\R$ and phase $\psi$ of the first harmonic 
is
\begin{equation}
\R = (a^2 + b^2)^{1/2}, \qquad \psi = \atan \Biggl(
{b\over a} \Biggr).
\end{equation}
Once we have a set of experimental UHECRs with amplitude $\R$ and phase
$\psi$, the probability $P(\S,\vartheta)$ that this data came from a
population characterized with an amplitude in $(\S,\S+d\S)$ and phase
$(\vartheta,\vartheta+d\vartheta)$ is given by [\cite{l75}]
\begin{equation}
P(\S,\vartheta)=\frac{\sqrt{N}}{2 \pi^{3/2} I_0(k_0/2)} \exp \Biggl[-
{N \over 4} \Bigl( \S^2+ \ffrac{1}{2}\R ^2 -2 \R \S \cos (\vartheta -\psi)
\Bigr) \Biggr],
\end{equation}
where $k_0 = \ffrac{1}{4}{N \R^2}$ and $I_0$ denotes the zeroth order
Bessel function of the first kind. Defining $\xi = \S/\R$, then the
differential probability distributions $P(\xi)$ and $P(\vartheta)$ are
obtained by straightforward integration as
\begin{eqnarray}
\label{eq:distsa}
P(\xi) &=& {2 k_0^{1/2} \over \pi^{1/2} I_0(k_0/2)} \exp \Bigl[
-{k_0\over 2} (1 + 2\xi^2) \Bigr] 
I_0 \Bigl( {2k_0\xi} \Bigr),\\
\label{eq:distsb}
P(\vartheta) &=& {1 \over 2 \pi I_0 (k_0/2)}
\Bigl[1 \pm \erf \left[ \pm k_0^{1/2} \cos(\vartheta -\psi)\right] \Bigr]
\exp \Bigl[ k_0(\cos^2 (\vartheta-\psi) -\ffrac{1}{2}) \Bigr].
\end{eqnarray}
Figure~\ref{fig:distributions} shows these probability distributions.
For large numbers of events, these distributions tend towards
normal. It is not obvious that the assumption of normal
distributions is appropriate for the AGASA and Haverah Park datasets,
as the number of events is rather modest.  In this paper, we always
calculate the dispersions about the expected value corresponding to a
given confidence limit by solving the integral equation over the
distribution. For example, the dispersion $\sigma_\vartheta$
corresponding to the $95 \%$ confidence limit is given by
\begin{equation}
\label{eq:confidence}
\int_{\langle \vartheta \rangle - \sigma_\vartheta}^{\langle \vartheta
\rangle + \sigma_\vartheta}
P(\vartheta) d\vartheta = 0.95.
\end{equation}
The symmetric dispersion on the amplitude $\sigma_\S$ can be defined
in the same way. However, the confidence limits on the amplitude $\S$
are asymmetric about the average values. So, sometimes it is useful
to use asymmetric dispersions $\sigma_\S^{+}$ and $\sigma_\S^{-}$ 
corresponding to $95 \%$ confidence limits defined via
\begin{equation}
\label{eq:asymconfidencea}
\int_{\langle \xi \rangle}^{\langle \xi
\rangle + \sigma_\xi^{+}} P(\xi) d\xi = 0.475,
\end{equation}
and
\begin{equation}
\label{eq:asymconfidenceb}
\int_{\langle \xi \rangle - \sigma_\xi^{-}}^{\langle \xi \rangle} 
P(\xi) d\xi = 0.475.
\end{equation}
Unless stated to the contrary, we use the asymmetric definitions
of confidence limits for the amplitude. This distinction can be
important at small $k_0$.

\subsection{The Data}

Table~\ref{table:tab1} summarises the available experimental data. The
Haverah Park shower array detected 27 UHECRs with energies above $4
\times 10^{19}$ eV and zenith angle less than $45^\circ$
[\cite{stanev}].  For the AGASA experiment, we use the data on the 57
UHECRs detected before May 2000 [\cite{agasa2}].
Table~\ref{table:tab1} records the amplitude $\R$ and phase $\psi$ as
determined from the two datasets\footnote{Since galactic magnetic
fields may influence the trajectories of cosmic rays with energies $<
4 \times 10^{19}$ eV [\cite{ptuskin}], we do not consider these events
in the analysis.}. The $95 \%$ confidence limits on the amplitude and
phase are given, from which it is already evident that {\it the
presently available data are too sparse to claim a secure detection of
any anisotropy}. This is in accord with AGASA's own analysis of the 47
events they observed before August 1998, in which they assert that
there is no evidence for significant large-scale anisotropy on the
celestial sphere.

Sommers~[\cite{ps}] has estimated that the likely size of the dataset
after 5 years of operation of the Auger Observatory is $\sim 2200$
events. This though is the combined total from both North and South
Auger Observatories. In this paper, we are interested in the prospects
over the next few years and so we restrict our analysis to South
Auger.  To give a feel for what it may achieve should there be a real
anisotropy in the UHECR sky, we present two further examples in
Table~\ref{table:tab1}. In the first (henceforth Auger I), we assume
that its amplitude and phase are the same as that indicated by AGASA,
but that the number of events detected over $\sim 4 \times 10^{19}$ eV
is $1000$.  In the second (Auger II), we assume the amplitude and
phase are the same as that indicated by Haverah Park (which uses
water-Cerenkov detectors of the same depth as Auger) and the number of
post-GZK events is $1000$. In both cases, the size of even the $95 \%$
confidence limits is reduced to typically $\sim 10^\circ$ in phase and
$\sim 30 \%$ in amplitude.

Table~\ref{table:tests} gives the details of further tests for
isotropy. We list the probability that an amplitude higher than that
measured $\R$ can arise from an underlying isotropic
distribution. This is given by the formula [\cite{l75}]
\begin{equation}
P ( > \R) = \exp (-k_0 ).
\end{equation}
We also give the probability that the data comes from a model with an
amplitude $\S$ between 0 (isotropic) and the symmetric dispersion
$\sigma_\S$ corresponding to the $68 \%$ and $95 \%$ confidence
limits. This table underlines that the present AGASA and Haverah Park
datasets are inadequate for the unambiguous identification of any
anisotropy signal, if it exists. For example, an amplitude equal to or
exceeding that reported by AGASA can come from an isotropic
distribution with a probability exceeding a third, which is hardly
negligible. The final two rows of Table~\ref{table:tests} show the
impact that South Auger will make. The small numbers for the
probabilities demonstrate that genuine anisotropy signals with the
same amplitude and phase as indicated by AGASA and Haverah Park will
be convincingly detected.


%
\begin{table}
\begin{center}
(a) Isothermal (Extent of 250 kpc)
\end{center}
\begin{center}
\begin{tabular}{|c|c|c|}\hline
Field Size &$ \Phi_{\rm M31} / \Phi_{\rm MW}$
           & $\Phi_{\rm M31} / \Phi_{\rm MW} $ \\ 
\null      & (Approx, eq.~\ref{eq:medtanc})
           & (Exact, eq.~\ref{eq:ff}) \\
\hline 
$2^\circ \times 2^\circ$  &39.7 &2.5 \\ \hline
$10^\circ \times 10^\circ$ &1.6 &0.6 \\ \hline
$15^\circ \times 15^\circ$ &0.7 &0.4 \\ \hline
\end{tabular}
\end{center}
\begin{center}
(b) Isothermal (Extent of 100 kpc)
\end{center}
\begin{center}
\begin{tabular}{|c|c|c|}\hline
Field Size &$ \Phi_{\rm M31} / \Phi_{\rm MW}$
           & $\Phi_{\rm M31} / \Phi_{\rm MW} $ \\ 
\null      & (Approx, eq.~\ref{eq:medtanc})
           & (Exact, eq.~\ref{eq:ff}) \\
\hline 
$2^\circ \times 2^\circ$  &16.4 &2.6 \\ \hline
$10^\circ \times 10^\circ$ &0.7 &0.5 \\ \hline
$15^\circ \times 15^\circ$ &0.3 &0.3 \\ \hline
\end{tabular}
\end{center}
\begin{center}
(c) Cusped (Extent of 250 kpc)
\end{center}
\begin{center}
\begin{tabular}{|c|c|c|}\hline
Field Size &$ \Phi_{\rm M31} / \Phi_{\rm MW}$
           & $\Phi_{\rm M31} / \Phi_{\rm MW} $ \\ 
\null      & (Approx, eq.~\ref{eq:medtanc})
           & (Exact, eq.~\ref{eq:ff}) \\
\hline 
$2^\circ \times 2^\circ$  &18.1 &3.0 \\ \hline
$10^\circ \times 10^\circ$ &0.7 &0.4 \\ \hline
$15^\circ \times 15^\circ$ &0.3 &0.2 \\ \hline
\end{tabular}
\end{center}
\begin{center}
(d) Cusped (Extent of 100 kpc)
\end{center}
\begin{center}
\begin{tabular}{|c|c|c|}\hline
Field Size &$ \Phi_{\rm M31} / \Phi_{\rm MW}$
           & $\Phi_{\rm M31} / \Phi_{\rm MW} $ \\ 
\null      & (Approx, eq.~\ref{eq:medtanc})
           & (Exact, eq.~\ref{eq:ff}) \\
\hline 
$2^\circ \times 2^\circ$  &11.8 &3.0 \\ \hline
$10^\circ \times 10^\circ$ &0.5 &0.4 \\ \hline
$15^\circ \times 15^\circ$ &0.2 &0.2 \\ \hline
\end{tabular}
\end{center}
\caption{Ratio between incoming UHECR flux from the M31 halo to that
from the Galaxy for fields of view of various sizes centered on M31.
The second column shows the dangers of the point mass approximation,
which overestimates the importance of the effect, while the third
column gives the exact result. In tables (a) and (b), both
galaxies are modelled with isothermal haloes, but with an extent 250
kpc and 100 kpc respectively. Tables (c) and (d) are the same but for
the use of the cusped halo models, rather than isothermals.}
\label{table:andromedaresults}
\end{table}

\section{The Dark Halo of the Galaxy}

\subsection{Models}

There is little definite evidence on the structure of the Galaxy's
dark matter halo. The HI gas rotation curve of the Milky Way cannot be
traced much beyond $\sim20$ kpc.  It is the kinematics of the distant
satellite galaxies that provide the best evidence on the total mass
and the extent of the Galaxy. The most recent estimate [\cite{mark}]
finds a total mass $\sim 2 \times10^{12}$M$_{\odot}$ and an extent of
$\gtrsim200$~kpc. Such a large extent is in accord with earlier
studies of the Milky Way and external spiral galaxies
[\cite{zaritsky}].

N-body computer simulations of structure formation in hierarchical
merging cosmogonies have provided evidence for a {\it cusped} density
profile for dark haloes [\cite{nfw96}]
\begin{equation}
n (\br) \propto \frac{1}{(r + \repsilon)(r + \rs)^2},
\label{eq:nfw}
\end{equation}
where $\br$ is the position with respect to the Galactic Centre and
$r=|\br|$ is the spherical polar radius. Here, $\rs$ is the scale
radius, which is typically $\sim10$~kpc for the Galaxy, while
$\repsilon$ is set by the resolution limit of the simulations,
nowadays $\sim0.5$~kpc. These models (hereafter NFW) have been much in
vogue over the last few years. For example, they have already been
used in calculations of the flux of UHECRs and the strength of the
anisotropy signal [\cite{bsw99,mtw99}]. The cusped density
distribution of the NFW profile necessarily implies that the inner
regions are dominated by dark matter and gives a substantial
anisotropy signal. However, there are three reasons why
the NFW is certainly incorrect for the Galaxy [\cite{york}]; first,
the mass density implied by the luminous disk and bar is already
sufficient to account for the rotation curve in the inner Galaxy
without any contribution from dark matter [\cite{ortwin}]; second,
numerical simulations of barred galaxies shows that copious amounts of
dark matter in the inner parts slow down and rapidly dissolve bars
through dynamical friction [\cite{victor}]; and third, the
microlensing optical depth to the red clump stars already shows that
almost all the density in the inner Galaxy must be in the form of
compact objects capable of causing microlensing and so cannot be
particle dark matter [\cite{jjb}].

It is more likely that the Galaxy's dark halo has an {\it isothermal}
distribution with a large core radius. Unlike the NFW profile, this at
least satisfies one of the few pieces of secure evidence --- namely
that luminous matter dominates the central regions, whereas dark
matter dominates the outer parts.  As a prototype of this model, we
take the cored isothermal-like sphere [\cite{nwe}]
\begin{equation}
 n(\br) \propto \frac{3\rc^2 + r^2}{(\rc^2 + r^2)^2}.
\label{eq:isothermal}
\end{equation}
Here, $\rc$ is the core radius, which is chosen to be $\sim10$~kpc.
This parameter is not constrained at all by the observations, but its
precise value does not have a big effect on the anisotropy of the
UHECRs. The model is the spherical limit of a more general family of
{\it triaxial} density distributions [\cite{wyn}]
\begin{equation}
n(\br)\propto \frac{Ax^2\!+\!By^2\!+\!Cz^2\!+\!D}
{(\rc^2\!+\!x^2\!+\!y^2p^{-2}\!+\!z^2 q^{-2})^2},
\label{eq:triaxial}
\end{equation}
where $\br = (x,y,z)$ and the constants $A,B,C$ and $D$ are
\begin{equation}
A=(p^{-2}\!+\!q^{-2}\!-\!1),\quad B= p^{-2}(1\!-\!p^{-2}\!+\!q^{-2}),
\quad C=q^{-2} (1\!+\!p^{-2}\!-\!q^{-2}),\quad D =
\rc^2(1\!+\!p^{-2}\!+\!q^{-2}).
\end{equation}
Here, $p$ and $q$ are axis ratios of the potential. When $p=1$, the
halo is oblate, whilst when $p=q$, the halo is prolate. The shape of
the Galaxy's dark halo is uncertain. The flaring of the outer HI gas
disk does provide direct evidence on the vertical force at the
midplane from the halo, which is greater for flattened models.
Analyses of this effect have been used to determine the shape of the
halo [\cite{OM}], but they are hampered by the unknown contribution to
pressure support from magnetic fields and cosmic rays at large
distances from the Sun.  There is also evidence on the shape of the
halo from stellar kinematics [\cite{rpvdm}], although this is beset
with modelling uncertainties as to the orientation of the velocity
ellipsoid. Such evidence as there is suggests a flattened dark halo
with an ellipticity less than E7.  However, even though evidence on
the shape of the dark halo is sparse, it is natural to expect halos
built from merging and accretion to possess flattened, generically
triaxial, shapes rather than spherical ones [\cite{warren}]. For the
sake of definiteness, we take $p=0.9$ and $q=0.75$, which means that
the density contours have semi-axes in the ratio $1:0.788:0.428$. This
is a highly flattened model with an ellipticity~\footnote{Recall that
the ellipticity is $10 \times (1 - b/a)$, where $b$ and $a$ are the
projected minor and major axes respectively. A spherical galaxy is E0,
while the most flattened elliptical galaxies are E7.} of roughly E6.
At first glance, a triaxial halo may seem to afford a great deal
of freedom as to the location of the Sun. This is of course not the
case! The Sun must be located close to one of the principal axes, as
otherwise there would be gross differences between the Galaxy as viewed
at positive and negative longitudes. Both local and global dynamical
evidence favour the location of the Sun as close to (within $10^\circ$
of) the major axis of the Galactic disk~[\cite{kt}].

There is one other effect that may be common. It has been emphasised
[\cite{james}] that the outer parts of the dark halo may be misaligned
with the disk. This is one possible cause of the phenomenon of warping
of the neutral gas disk, an effect that is known to be present in the
outer Milky Way. A family of {\it tilted} halo models has density
\begin{equation}
n(\Rp,\zp)\propto \frac{\rc^2 (2\!+\!q^{-2})\!+\!q^{-2}\Rp^2\!
 +\!\zp^2 q^{-2}(2\!-\!q^{-2})}{(\rc^2\!+\!\Rp^2\!+\!\zp^2 q^{-2})^2},
\label{eq:tilted}
\end{equation}
where $\Rp^2 = \xp^2 + \yp^2$. The coordinates ($\xp,\yp,\zp$) are
related to ($x,y,z$) by a rotation through an angle $\theta$ about the
$x$-axis, on which the Sun lies. The motivation for this is that the
Sun lies nearly on the line of nodes of the warp. We take $\theta =
30^\circ$, an extreme value, as we are interested in the qualitative
effects of tilted haloes.

There are two astrophysical parameters we will need to calculate
signatures of the UHECRs, assuming they arise from the decay of heavy
relics in the halo. The first is $\rsun$, the distance from the Sun to
the Galactic Centre, which is taken as $(8.5,0,0)$ [\cite{kerr}]. The
second is $\rmax$, the extent of the halo, which is taken as $\sim
250$~kpc in accord with the most recent estimates
[\cite{mark,zaritsky}]. The halo extent is poorly known and results
depending sensitively on this parameter must be viewed with some
caution.

\begin{figure}
\begin{center}
\begin{tabular}{c@{\hspace{2cm}}c}
  \psfrag{r}{$\S$}
  \psfrag{p}{$\vartheta$}
  \includegraphics[width=7cm,height=7cm]{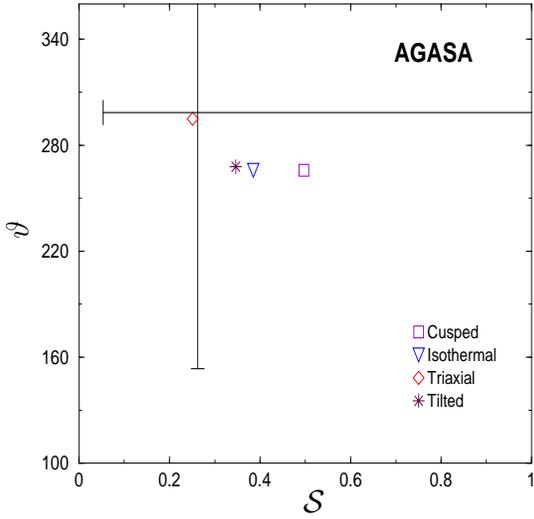}&
  \psfrag{r}{$\S$}
  \psfrag{p}{$\vartheta$}
  \includegraphics[width=7cm,height=7cm]{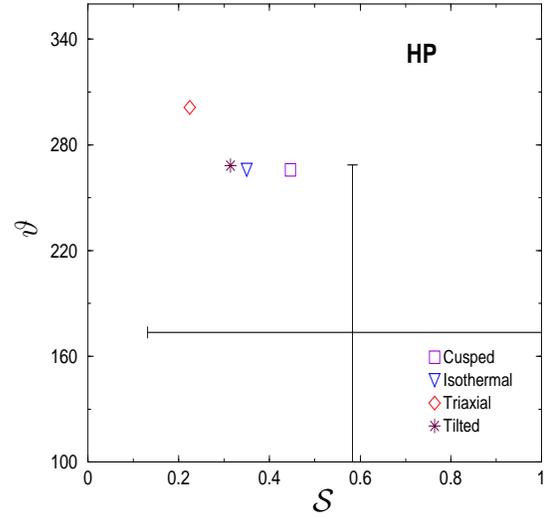}\\
(a) & (b) \\
&\\
&\\
  \psfrag{r}{$\S$}
  \psfrag{p}{$\vartheta$}
  \includegraphics[width=7cm,height=7cm]{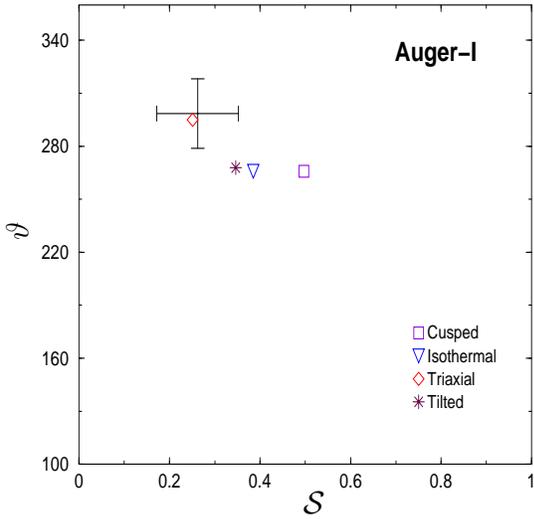}&
  \psfrag{r}{$\S$}
  \psfrag{p}{$\vartheta$}
  \includegraphics[width=7cm,height=7cm]{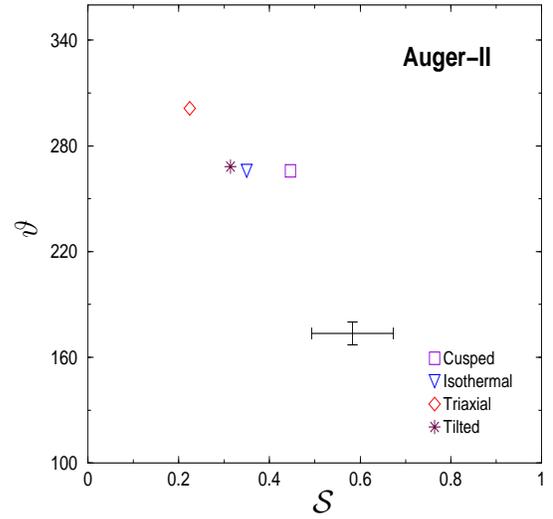}\\
(c) & (d) 
\end{tabular}
\end{center}
\caption{Plots of the amplitude against the phase for the four models
(cusped, isothermal, triaxial and tilted). Also shown are the $95 \%$
confidence limits for (a) the AGASA and (b) the Haverah Park
experiments. The lower panels show the expected impact of South
Auger. Marked are the $95 \%$ confidence limits assuming the amplitude
and phase of (c) the AGASA and (d) the Haverah park experiments, but
with 1000 events.}
\label{fig:harmonicfig}
\end{figure}
%

\subsection{Results}

Figure~\ref{fig:allsky} shows what our four halo models look like in
the cosmic ray sky. These are contours of the incoming UHECR flux in
equatorial coordinates.  Figure~\ref{fig:allskyefficiency} shows the
contours convolved with the response function $h_1(\delta)$ for South
Auger. (The figure for $h_2(\delta)$ is very similar).  If the UHECRs
do originate in dark haloes, then the halo of the nearby Andromeda
galaxy (M31) may give a tell-tale signature. M31 is the only other
large galaxy in the Local Group, and it is roughly as massive as the
Galaxy [\cite{ewggv}].  The halo of M31 is modelled as an isothermal
sphere centered 770 kpc away and with an extent of $250$ kpc. In
Figures~\ref{fig:allsky} and \ref{fig:allskyefficiency}, the hotspot
at $\alpha_0 \approx 00^{\rm h} 43^{\rm m}, \delta_0 \approx 41^\circ$
(referred to the J2000.0 epoch) is caused by M31. It has already been
claimed that the absence of a hotspot in the data rules out the
decaying dark matter origin of UHECRs [\cite{bsw99}], though this has
been contested by others [\cite{mtw99}].

The issue is worth considering in more detail.  The ratio of the mass
of M31 to that of the Galaxy used previously [\cite{bsw99,mtw99}] is
two to one. There is little dynamical evidence for such a ratio.  The
asymptotic rotation curve of M31 is just $\sim 10 \%$ higher than that
of the Galaxy.  The most reliable estimate for the mass of the M31
halo based on the motions of satellite galaxies, planetary nebulae and
globular clusters is $\sim 1 \times 10^{12}\msun$, though there is at
least a factor of two uncertainty in this value due to the small sizes
of the tracer datasets [\cite{ewggv}]. On balance, it is probably more
accurate to assume that M31 is roughly as massive as the Galaxy.  Now,
to compute the ratio of the total flux received from M31 to that
received from the Milky Way $\Phi_{\rm M31}/\Phi_{\rm MW}$,
the following expression has sometimes been used [\cite{mtw99}]:
\begin{equation}
\frac{ \Phi_{\rm M31}}{\Phi_{\rm MW}} \sim \frac{\xi}{D^2} \times
  \frac{\int_{V_{\rm M31}} n_{\rm M31}\, dV}{\int_{V_{\delta \Omega}} n_{\rm
  MW}/r^2\, dV}.
\label{eq:medtanc}
\end{equation}
Here, $\xi$ is the ratio between the masses of the two haloes, $D \sim
770$ kpc, $V_{\rm M31}$ is the volume of the halo of M31 and
$V_{\delta \Omega}$ is the volume defined by the cone of solid angle
$\delta \Omega$ pointing towards M31.  There are two drawbacks to this
formula. First, it assumes that M31 is point-like so that $n_{\rm
M31}/r^2$ is taken as constant and equal to its value at the center of
M31. Second, it assumes that the entire halo of M31 fits into the
angular window. Accordingly, we prefer to use the exact expression
\begin{equation}
\frac{ \Phi_{\rm M31}}{\Phi_{\rm MW}} \sim 
\frac{\int_{\alpha_0-\Delta\alpha/2}^{\alpha_0+\Delta\alpha/2}\, d\alpha
\int_{\delta_0-\Delta\delta/2}^{\delta_0+\Delta\delta/2} \cos
\delta\,d \delta
\int_{s_-}^{s_+}  n_{\rm M31}\,ds}
{\int_{\alpha_0-\Delta\alpha/2}^{\alpha_0+\Delta\alpha/2}\,d\alpha
\int_{\delta_0-\Delta\delta/2}^{\delta_0+\Delta\delta/2} \cos \delta
\,d\delta
\int_{0}^{s_{\rm max}}n_{\rm MW}\,ds}.
\label{eq:ff}
\end{equation}
Here, the flux ratio is measured in an angular window $\Delta \alpha
\times \Delta \delta$ centered on the right ascension $\alpha_0$ and
declination $\delta_0$ of M31. The quantities $s_\pm$ are computed by
finding the intersection of the line of sight towards with a sphere of
radius $250$ kpc, while $s_{\rm max}$ is the usual limit of the Galaxy
halo.  Both the Galaxy and M31 are modelled with identical cored
isothermal haloes.  Table~\ref{table:andromedaresults} gives the
results for the flux ratio computed using the approximate
formula~(\ref{eq:medtanc}) and the exact one~(\ref{eq:ff}).
Results are presented when both the Galaxy and M31 are modelled
with isothermal and cusped haloes of extents of 250 kpc and 100 kpc.
We see that the effects of an extended dark halo are not always
well-represented by the point-like approximation. However,
formula~(\ref{eq:medtanc}) is still valuable, as it gives an upper
limit to M31's contribution.  Given that the expected spread caused
by the deflection of a $4 \times 10^{19}$ eV proton originating from
Andromeda is of the order of a few degrees, a field of view of
$10^\circ \times 10^\circ$ is probably the most appropriate one to
look for the enhancement effect of the M31 halo [\cite{mtw99}].  On
such scales, no substantial enhancement effect is expected and the
dominant source of UHECRs remains the Galaxy.We therefore agree
with the assessment of Medina Tanco \& Watson [\cite{mtw99}] -- and
disagree with Benson et al. [\cite{bsw99}] -- as regards the
visibility of M31 in the UHECR sky.

%
\begin{table}
\begin{center}
{\bf (a) AGASA}
\end{center}
\begin{center}
\begin{tabular}{|c|c|c|c|c|c|} \hline
Model & Amplitude & Phase & $P (\sigma_\S, \sigma_\vartheta)$
& $P(\sigma_\S)$ & $P(\sigma_\vartheta)$ \\ 
\null & $\S$ & $\vartheta$ 
& $(95 \%)$ & $(95 \%)$ & $(95 \%)$  \\ \hline 
Cusped&0.497&265.8&0.328&0.330&0.944\\ \hline
Isothermal&0.385&265.8&0.563&0.572&0.944\\ \hline 
Triaxial&0.251&295.0&0.865&0.898&0.950\\ \hline
Tilted&0.346&267.8&0.651&0.665&0.945\\ \hline
\end{tabular}
\end{center}
\begin{center}
{\bf (b) Haverah Park}
\end{center}
\begin{center}
\begin{tabular}{|c|c|c|c|c|c|} \hline
Model & Amplitude & Phase & $P (\sigma_\S, \sigma_\vartheta)$
& $P(\sigma_\S)$ & $P(\sigma_\vartheta)$ \\ 
\null & $\S$ & $\vartheta$
& $(95 \%)$ & $(95 \%)$ & $(95 \%)$ \\ 
\hline 
Cusped&0.446&265.8&0.530&0.977&0.543\\ \hline
Isothermal&0.350&265.8&0.515&0.951&0.542\\ \hline 
Triaxial&0.224&301.2&0.172&0.887&0.180\\ \hline
Tilted&0.314&268.2&0.480&0.937&0.513\\ \hline
\end{tabular}
\end{center}
\caption{For each of the models, the theoretical values of the
amplitude and the phase are calculated using the response functions
for (a) AGASA and (b) Haverah Park. We also give the probabilities
that the experimental amplitude and phase come from within the $95 \%$
confidence limits of the model.}
\label{table:tab4}
\end{table}
%
%
\begin{table}
\begin{center}
{\bf (a) Auger I}
\end{center}
\begin{center}
\begin{tabular}{|c|c|c|c|c|c|c|} \hline
Model & $P(\sigma_\S, \sigma_\vartheta)$ & $P(\sigma_\S)$ 
& $P(\sigma_\vartheta)$\\ 
\null & ($95 \%$) & ($95 \%$) & ($95 \%$) \\ \hline 
Cusped&$<10^{-3}$&$<10^{-3}$&$0.095$\\ \hline
Isothermal&$0.014$&0.202&0.095\\ \hline 
Triaxial&0.886&0.950&0.932\\ \hline
Tilted&0.059&0.509&0.133\\ \hline
\end{tabular}
\end{center}
\begin{center}
{\bf (b) Auger II}
\end{center}
\begin{center}
\begin{tabular}{|c|c|c|c|} \hline
Model & $P(\sigma_\S, \sigma_\vartheta)$ & $P(\sigma_\S)$ 
& $P(\sigma_\vartheta)$\\ 
\null & ($95 \%$) & ($95 \%$) & ($95 \%$) \\ \hline 
Cusped    &$<10^{-3}$ &$0.156$ &$<10^{-3}$\\ \hline
Isothermal&$<10^{-3}$ &$0.001$ &$<10^{-3}$\\ \hline
\end{tabular}
\end{center}
\begin{center}
{\bf (c) Predictions for South Auger}
\end{center}
\begin{center}
\begin{tabular}{|c|c|c||c|c|}\hline
Model&Amplitude &Phase &Amplitude &Phase \\ 
\null& $(h_1)$ & $(h_1)$ & $(h_2)$& $(h_2)$\\ \hline
Cusped&0.883&265.8&0.806&265.8\\ \hline
Isothermal&0.558&265.8&0.508&265.8 \\ \hline
Triaxial&0.551&257.9&0.508&255.9 \\ \hline
Tilted&0.512&265.0&0.457&264.6\\ \hline
\end{tabular}
\end{center}
\caption{For each of the models, this gives an idea of what we will
learn from the southern station of the Auger Observatory. Table (a)
assumes the \ch{amplitude and phase} of the AGASA experiment, table
(b) the \ch{amplitude and phase} from Haverah Park; in both cases,
1000 events are analysed as typical of the number of UHECRs recorded
by South Auger. In Table (b), only the results for the isothermal and
cusped models are recorded; for the other models, all the
probabilities are less than $10^{-3}$. Table (c) presents the
predictions for our halo models using the two possible response
functions $h_1(\delta)$ and $h_2(\delta)$.}
\label{table:tab5}
\end{table}
%
%

Table~\ref{table:tab4} examines the four halo models (cusped,
isothermal, triaxial and tilted). For the moment, we assume the
canonical values of the halo parameters given in Section 3.1.  For
each model, we compute the theoretical amplitude $\S$ and phase
$\vartheta$ in the first two columns. Then, the next columns give the
probability that the experimental data came from a particular model
with $\S$ in $(\S-\sigma_\S,\S+\sigma_\S)$ and $\vartheta$ in
$(\vartheta-\sigma_\vartheta,\vartheta+\sigma_\vartheta)$.  This is
found by computing the integral
\begin{equation}
P(\sigma_S, \sigma_\vartheta) = \int_{\S-\sigma_\S^-}^{\S+\sigma_\S^+}
\int_{\vartheta-\sigma_\vartheta}^{\vartheta+\sigma_\vartheta}
{P(\S,\vartheta) d\S d\vartheta}.
\end{equation}
The final two columns give the equivalent results for the
one-dimensional distributions of amplitude and phase.

For both AGASA and Haverah Park response functions, the amplitude is
greatest for the cusped NFW model. This makes good sense, as this
model has the largest central density. We have argued above that the
Galaxy almost certainly does not have such a cusped density
distribution for its dark halo, and so a better guide to the expected
amplitude is provided by the other three halo models. Isothermal-like
models with a core radius $\sim 10$ kpc have an anisotropy of
amplitude $\S \lesssim 0.4$ as measured by the AGASA and Haverah Park
experiments.  For all four models, the phase points roughly towards
the Galactic Centre at $\alpha \approx 17^{\rm h} 45^{\rm m}, \delta
\approx -29^\circ$. The largest deviation from this direction of 
$35^\circ$ occurs for the triaxial halo. Notice that the probabilities
listed in the final three columns are all far from insignificant. The
data are compatible with any of the models. The present dataset is
insufficient to discriminate between possible dark matter halo
densities.

In Table~\ref{table:tab5} (a), which corresponds to Auger I, we see
the improvement in discrimination. The amplitude and phase is the same
as that seen by AGASA (see Table~\ref{table:tab1}) and happens to lie
close to the prediction of the triaxial model. Hence,
Table~\ref{table:tab5} (a) shows that all the probabilities are very
small, except that of the triaxial model. In other words, there is now
successful discrimination between halo models. All the probabilities
in Table~\ref{table:tab5} (b), which corresponds to Auger II, are very
small. The phase, assumed to be that of Haverah Park, does not lie
close to any of the halo models which are accordingly ruled out.  In
Table~\ref{table:tab5} (c), we give the expected amplitude and phase
for our four halo models and convolved with the two different response
functions $h_1(\delta)$ and $h_2(\delta)$ proposed for South
Auger. These results are shown pictorially in
Figure~\ref{fig:harmonicfig}.  The panels show the datasets from
AGASA, Haverah Park, Auger I and II, together with $95\%$ confidence
limits.

The anisotropy of all four halo models has a larger amplitude ($\S
\gtrsim 0.5$) at the South Auger detector ($35^\circ$ S) than at AGASA
($36^\circ$ N) or Haverah Park ($54^\circ$ N). For the spherical
models, the phase marks the direction of the Galactic Centre. A
signature of triaxiality is that the phase has different angular
offsets from the Galactic Centre direction at different detector
locations.

\begin{figure}
\begin{center}
\begin{tabular}{c@{\hspace{2cm}}c}
  \psfrag{r}{$\S$}
  \psfrag{p}{$\vartheta$}
  \psfrag{b}{$\scriptstyle r_{\rm s}=100\, {\rm kpc}$}
  \psfrag{s}{$\scriptstyle r_{\rm s}=1\, {\rm kpc}$}
  \includegraphics[width=6.5cm,height=6cm]{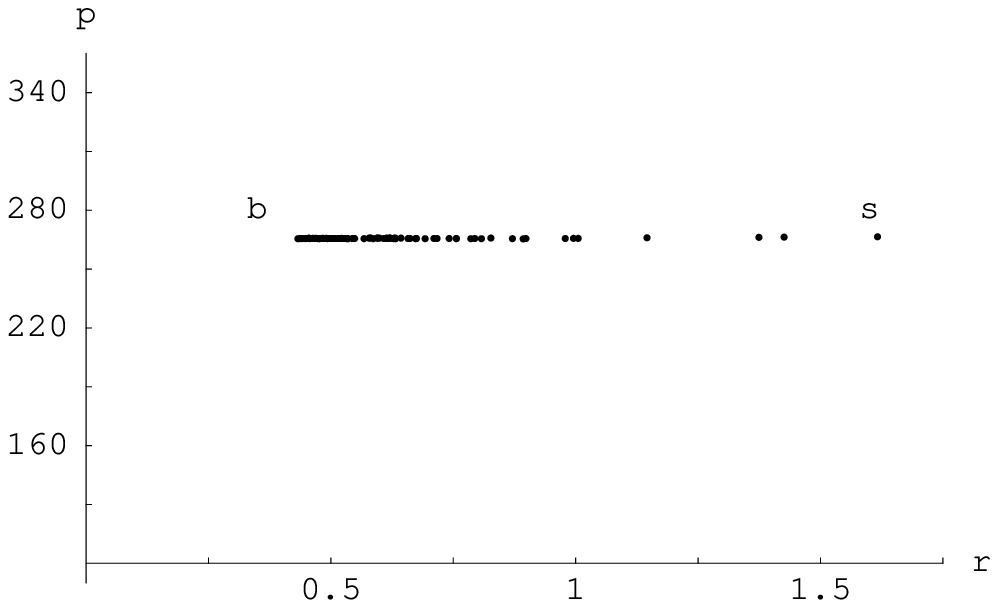}&
  \psfrag{r}{$\S$}
  \psfrag{p}{$\vartheta$}
  \psfrag{b}{$\scriptstyle r_{\rm c}=100\, {\rm kpc}$}
  \psfrag{s}{$\scriptstyle r_{\rm c}=1\, {\rm kpc}$}
  \includegraphics[width=6.5cm,height=6cm]{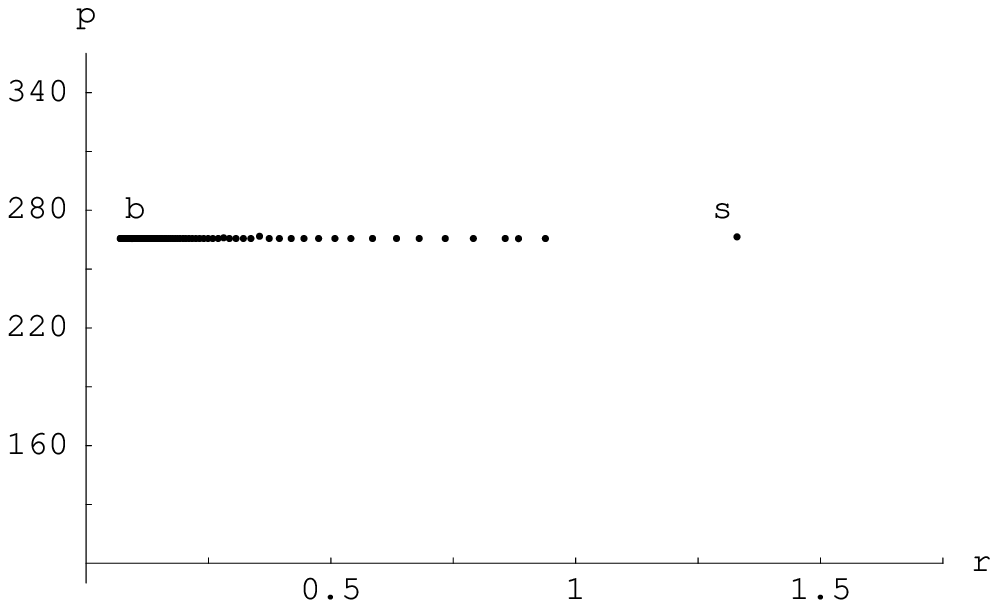}\\
(a) & (b) \\
&\\
&\\
  \psfrag{r}{$\S$}
  \psfrag{p}{$\vartheta$}
  \psfrag{h}{$\scriptstyle p=q$}
  \psfrag{l}{$\scriptstyle p=1$}
  \includegraphics[width=6.5cm,height=6cm]{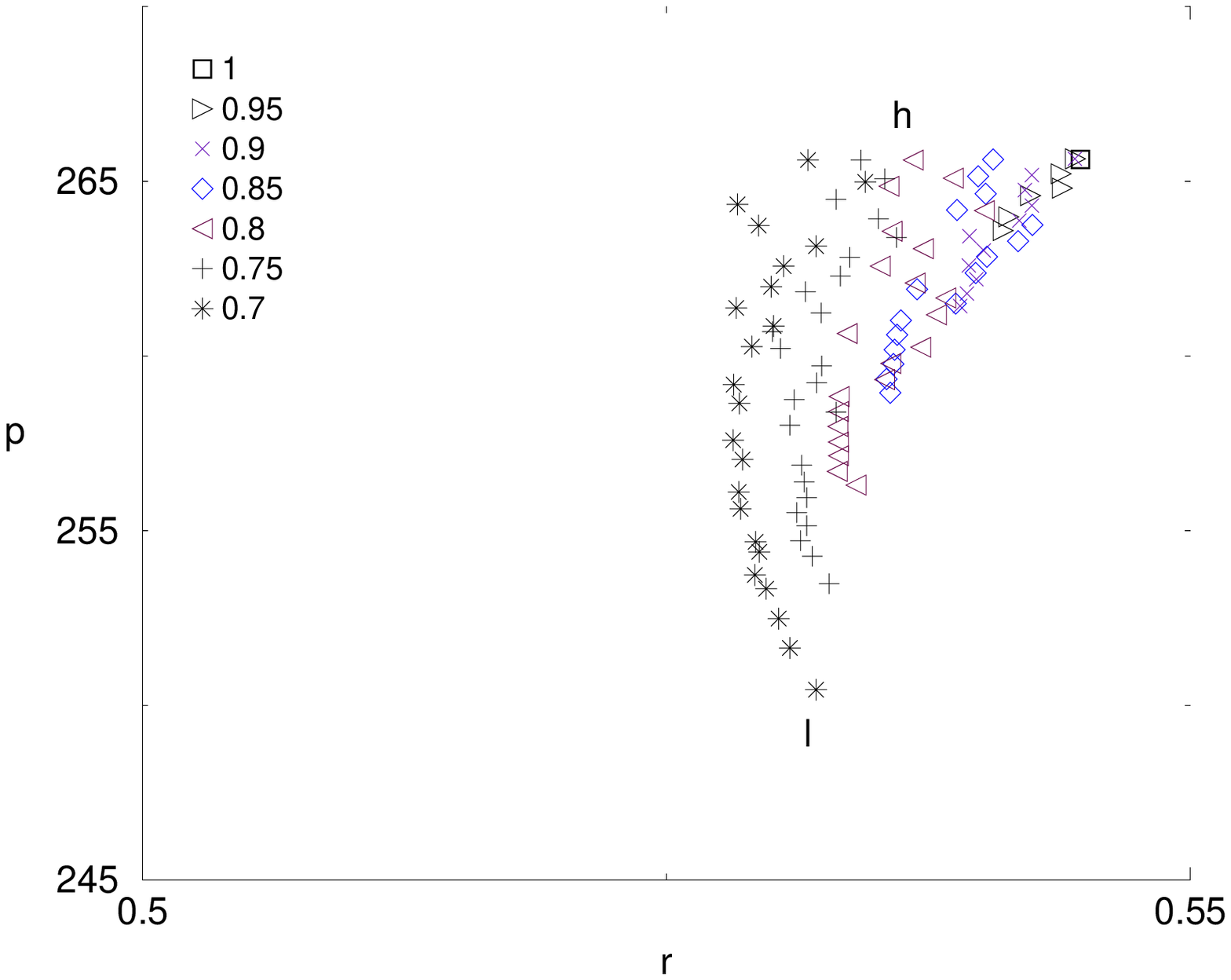}&
  \psfrag{r}{$\S$}
  \psfrag{p}{$\vartheta$}
  \psfrag{a}{$\scriptstyle \theta=40^\circ$}
  \psfrag{b}{$\scriptstyle \theta=-40^\circ$}
  \includegraphics[width=6.5cm,height=6cm]{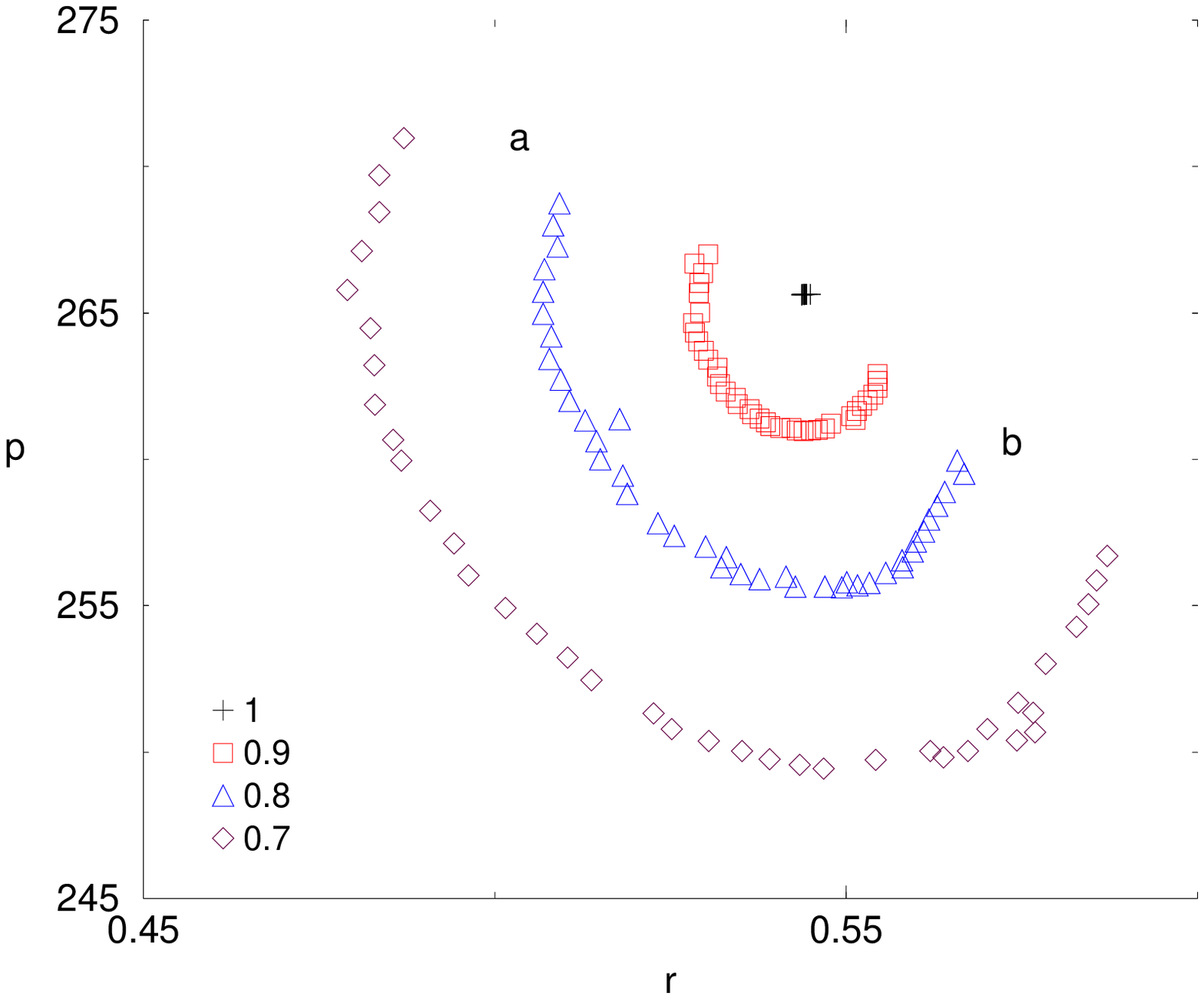}\\
(c) & (d)
\end{tabular}
\end{center}
\caption{Amplitude-phase plots showing how the anisotropy signal
measured by South Auger varies with the parameters in the (a) cusped,
(b) isothermal, (c) triaxial and (d) tilted halo models. The legend in
(c) and (d) corresponds to the flattening in the potential $q$.}
\label{fig:modelvariations}
\end{figure}

Figure~\ref{fig:modelvariations} investigates the amplitude $\S$ and
phase $\vartheta$ seen by South Auger as a function of the model
parameters for the four halo models. Panels (a) and (b) show the
effects of varying the scale radius $\rs$ of the cusped NFW halo and
the core radius $\rc$ of the isothermal halo. These parameters affect
only the amplitude of the harmonic and not the phase, which is always
in the direction of the Galactic Centre. Note that there is no simple
way of measuring the core radius of a dark halo from astrophysical
data. Usually, the core radius is inferred by least-square fitting to
the galaxy rotation curve, but this is an uncertain procedure as a
single rotation curve must be decomposed into separate contributions
from the bulge, disk and halo.  If UHECRs are indeed produced by
decaying dark matter, these plots raise the exciting possibility that
halo core radius may become directly accessible to measurement.
During the first 5 years of operation of South Auger, the core
radius could be measured if $\rc \lesssim 10$ kpc.  Panel (c) shows
the effect of varying the shape of the triaxial halo. The oblate
models ($p=1$) show largest deviation in phase ($\sim 15^\circ$) from
the spherical model ($p=q=1$). By comparison, stretching or flattening
the halo does not change the value of the amplitude $\S$ by much. In
panel (d), the tilt angle is varied. This can give large deviations in
phase ($\sim 25^\circ$) and significant changes in amplitude. Overall,
Figure~\ref{fig:modelvariations} suggests the shape of the halo
controls the phase $\vartheta$, whereas the scale of the halo controls
the amplitude $\S$.

\begin{figure}
\begin{center}
\begin{tabular}{c@{\hspace{2cm}}c}
  \psfrag{r}{$r_{\rm s}$}
  \psfrag{p}{$\!\!\!\!\!\!\!\!{\cal N}_\S$}
  \includegraphics[width=6.5cm,height=6cm]{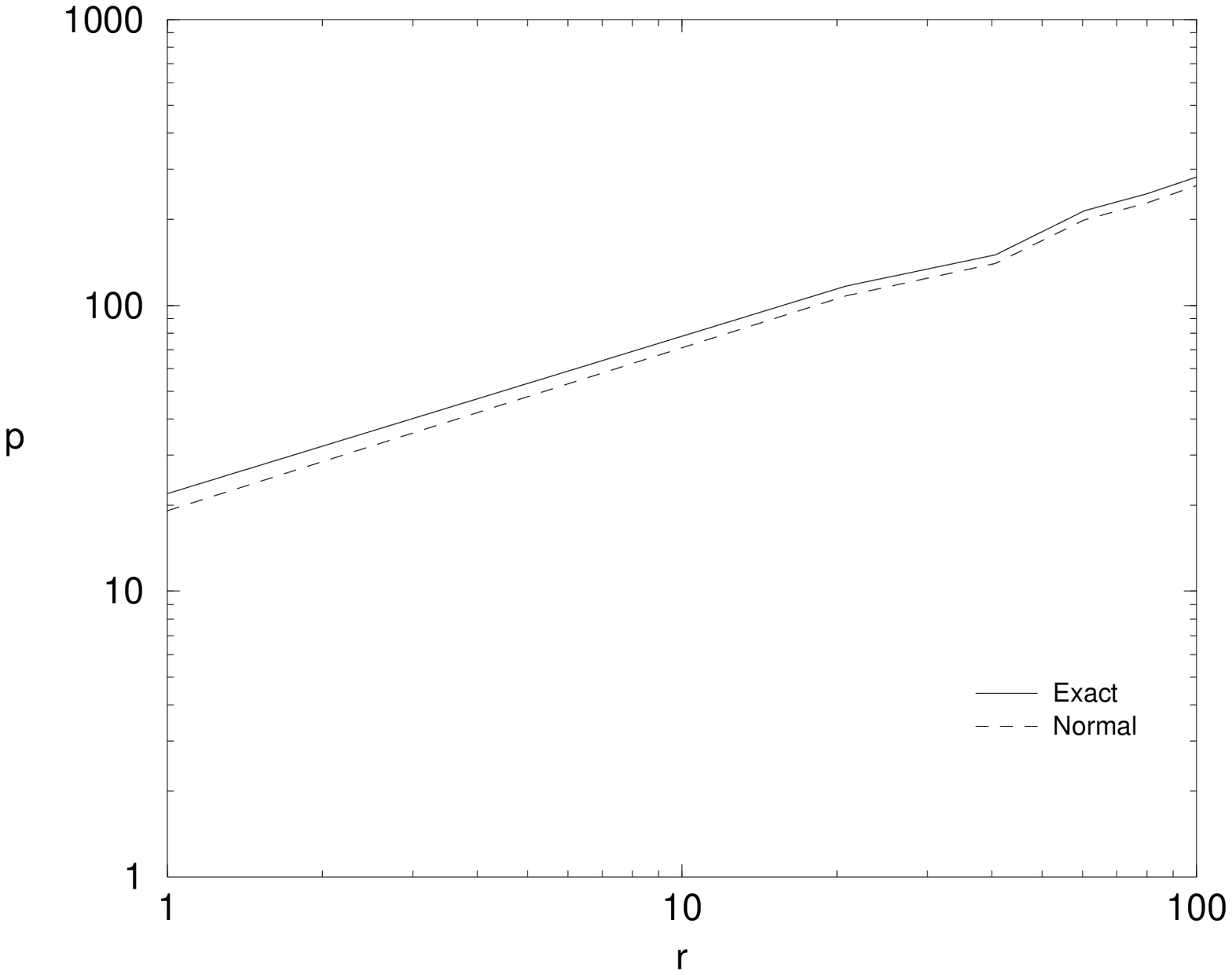}&
  \psfrag{r}{$r_{\rm c}$}
  \psfrag{p}{$\!\!\!\!\!\!\!\!{\cal N}_\S$}
  \includegraphics[width=6.5cm,height=6cm]{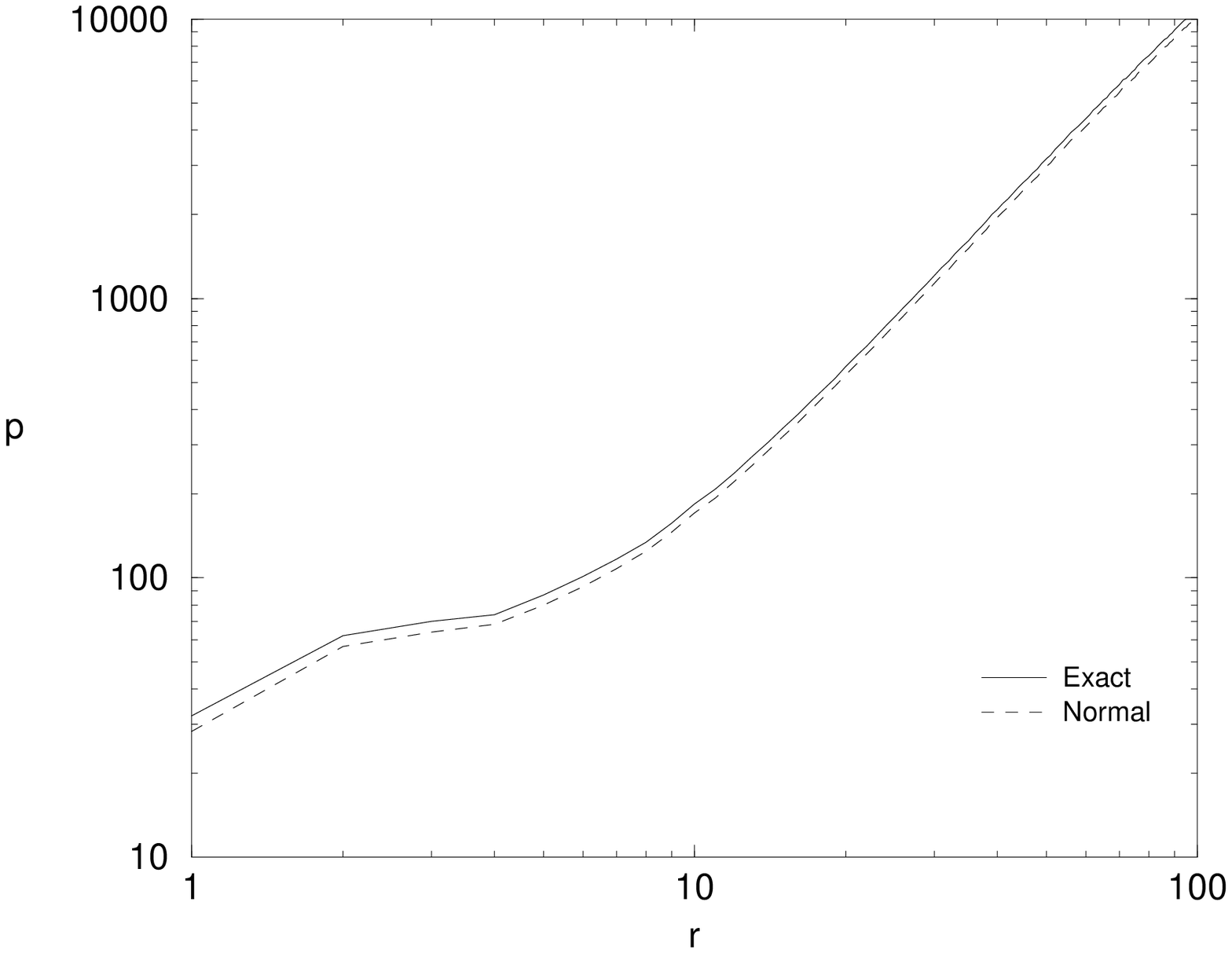}\\
(a) & (b) \\
&\\
&\\
  \psfrag{r}{$p$}
  \psfrag{p}{$\!\!\!\!\!\!\!\!{\cal N}_\S$}
  \includegraphics[width=6.5cm,height=6cm]{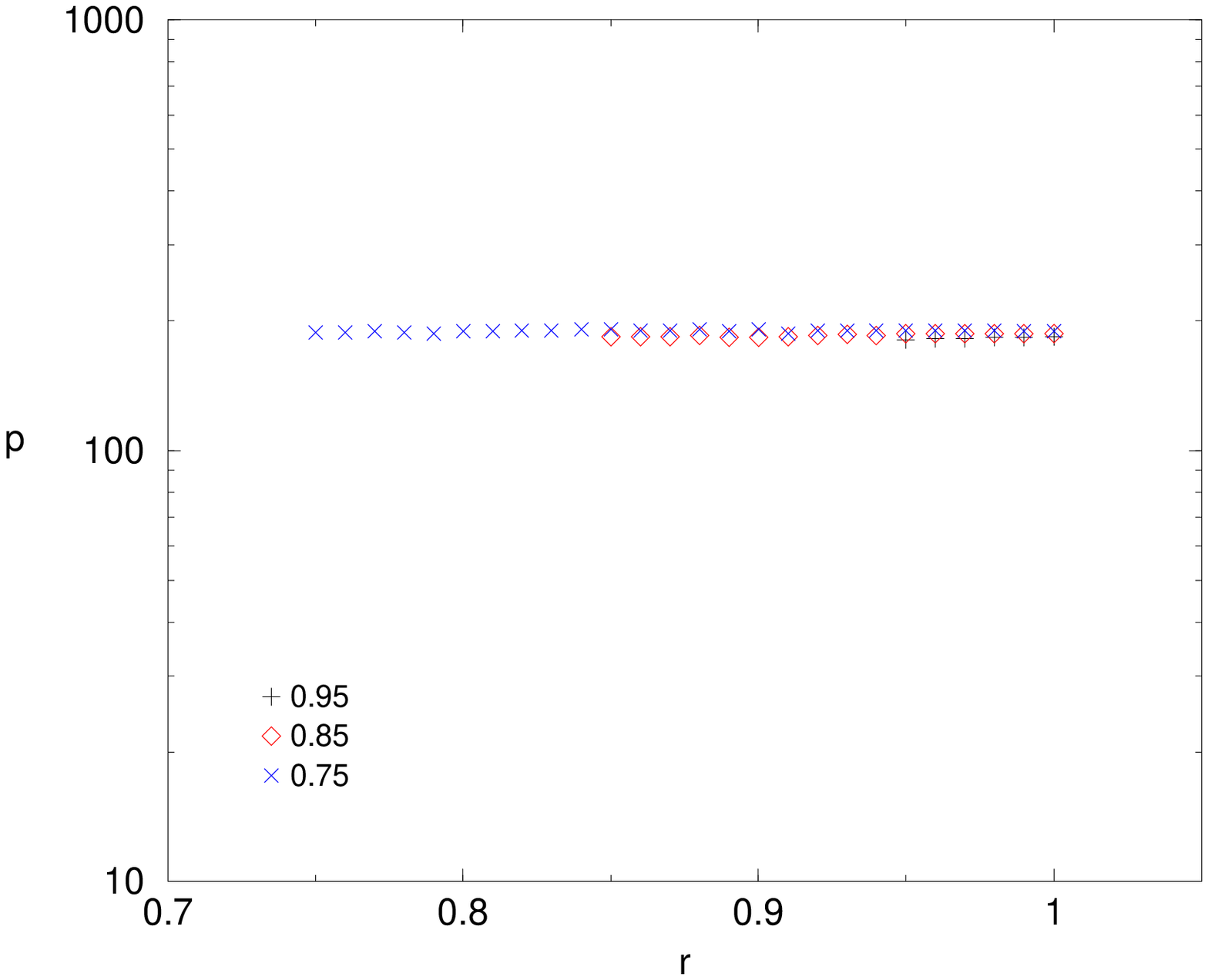}&
  \psfrag{r}{$\theta$}
  \psfrag{p}{$\!\!\!\!\!\!\!\!{\cal N}_\S$}
  \includegraphics[width=6.5cm,height=6cm]{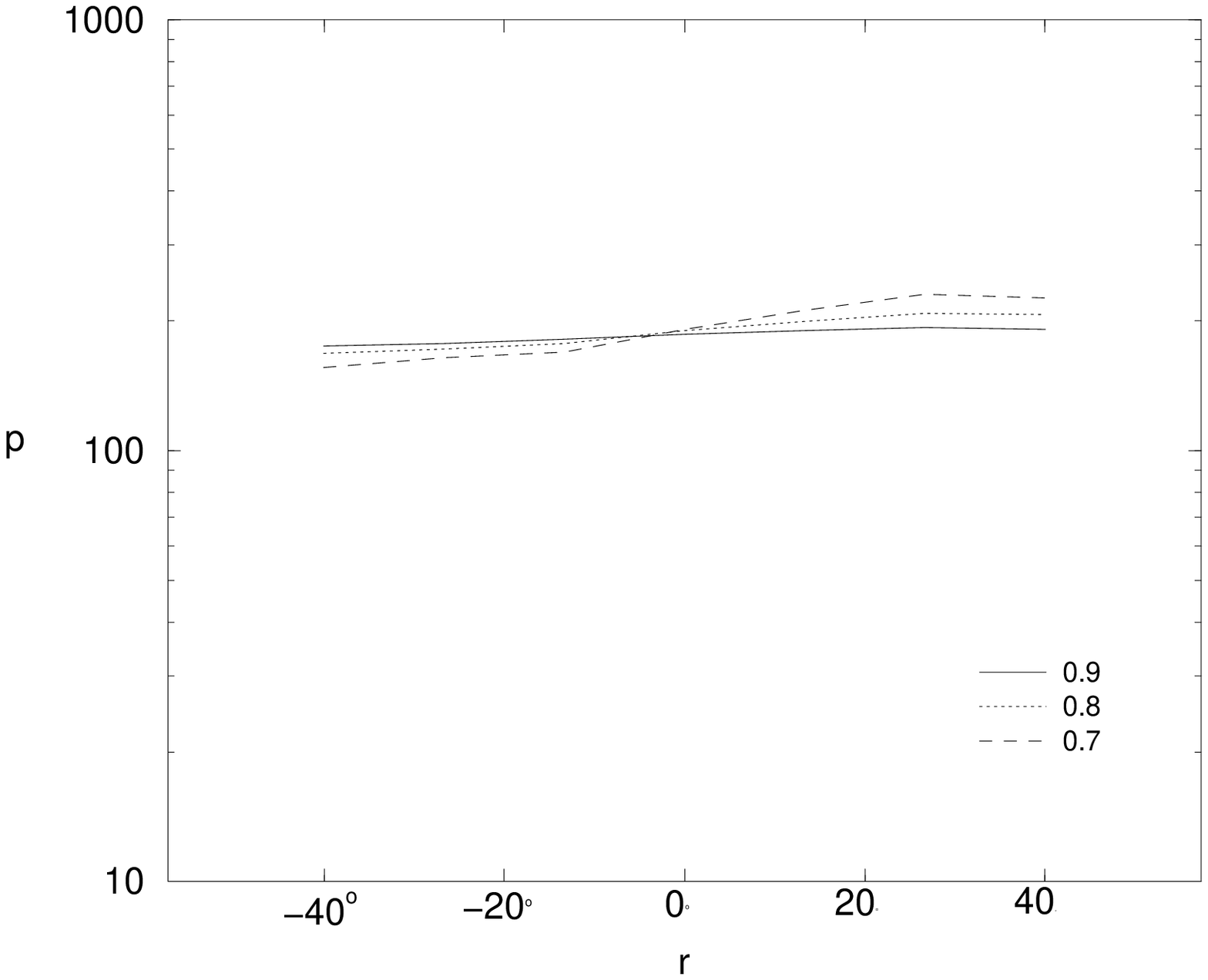}\\
(c) & (d)
\end{tabular}
\end{center}
\caption{Number of events required for 5$\sigma$ detection of an
anisotropy by South Auger for the (a) cusped, (b) isothermal, (c)
triaxial and (d) tilted halo models. The full and dashed curves in (a)
and (b) refer to the exact result and the normal approximation
respectively. The legend in (c) and (d) corresponds to the flattening
in the potential $q$.}
\label{fig:counts}
\end{figure}

Figure~\ref{fig:counts} shows the the number of events ${\cal N}_\S$
required for a convincing $5 \sigma$ detection of the anisotropy
signal for the four halo models at South Auger. An approximate method
is to use
\begin{equation}
{\cal N}_\S \approx {2 n_\sigma^2 \over \S^2}.
\label{eq:nsigmaruf}
\end{equation}
This gives the number of events required for a signal-to-noise ratio
of $n_\sigma$ standard deviations in amplitude. This assumes that the
probability density $P(\S)$ is a normal distribution about $\S = \R$
with dispersion equal to $\sqrt{2/{\cal N}_\S}$. This is not the case,
especially when the number of events is small or the amplitude is
small (i.e., $k_0$ is small). In an exact calculation, the
signal-to-noise is computed from its definition
\begin{equation}
n_\sigma \equiv \frac{\langle \S \rangle}{\sqrt{\sigma_\S^2}},
\label{eq:nsigma}
\end{equation}
where
\begin{displaymath}
\langle \S \rangle \equiv \int_0^\infty d\S\, \S P(\S),
\qquad\qquad
\sigma_\S^2 \equiv \left\{ \int_0^\infty d\, \S \S^2 P(\S) \right\} -
\langle \S \rangle^2.
\end{displaymath}
If we fix $n_\sigma=5$, as is required for a convincing detection,
then eq.~(\ref{eq:nsigma}) is an implicit equation for ${\cal N}_\S$,
which can be solved iteratively.  Figure~\ref{fig:counts} (a) and (b)
show the number of events for a $5\sigma$ detection calculated using
the exact expression (full curves) and the normal approximation
(dashed curves), from which it is evident that the approximation tends
to underestimate the required number of events.  Nonetheless,
given the ease with which the normal approximation can be computed, it
is reassuring to see that it is never seriously in error.  Roughly 40
events may be needed if the Galaxy's halo has a cusped NFW profile
with a small scale length $\rs < 2$ kpc. More likely, if the halo is
an isothermal sphere with large core radii ($\rc \sim 10-20$ kpc),
then typically 150-500 events may be required. Thus, within 5 years of
operation, South Auger will be able to rule some halo models out.

Panels (c) and (d) show that shape parameters like triaxiality and
tilt do not have a substantial influence on the number of required
events for a convincing measurement of the amplitude. For these
models, it is more interesting to ask how many events $N_\varphi$ are
needed for the error bars on the phase to be reduced to $\pm
\beta$. Using the normal approximation, this gives
\begin{equation}
{\cal N}_\varphi \approx {2 n_\sigma^2 \over \S^2}{1\over \sin^2 \beta}.
\end{equation}
For example, for the $3 \sigma$ error bars on the phase measured at
South Auger to be reduced to $\pm 10^\circ$ requires $\sim 2000$
events. This is the number needed to begin to distinguish between
different triaxial or tilted models and is possible after $\gtrsim 10$
years of operation. A less ambitious target is to ask for the $3
\sigma$ error bars to be $\pm 20^\circ$, which requires $\sim 500$
events. This is ample to verify that the phase points in the rough
direction of the Galactic Centre and thus implicate the decaying dark
matter models as the correct mechanism for UHECR production.

\begin{figure}
\begin{center}
\begin{tabular}{cc}
\includegraphics[width=7.5cm,height=4cm]{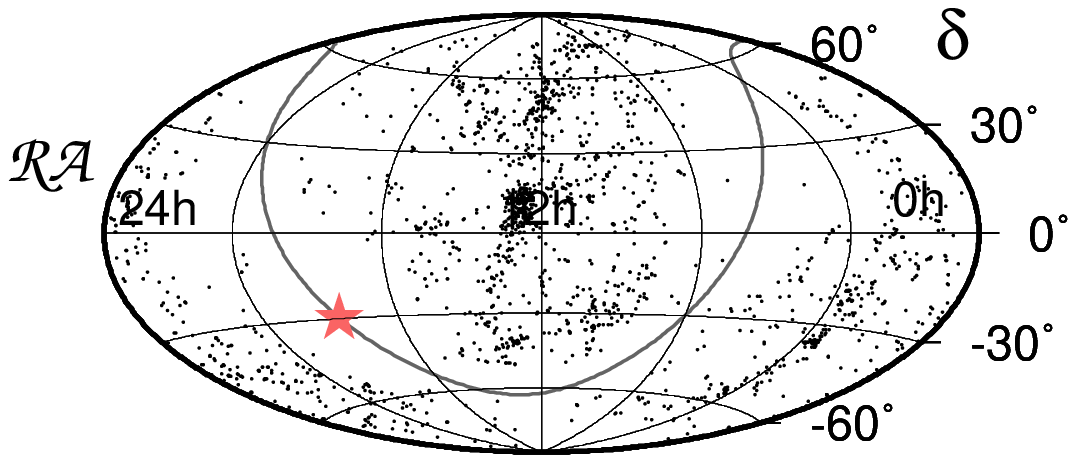}&
\includegraphics[width=7.5cm,height=4cm]{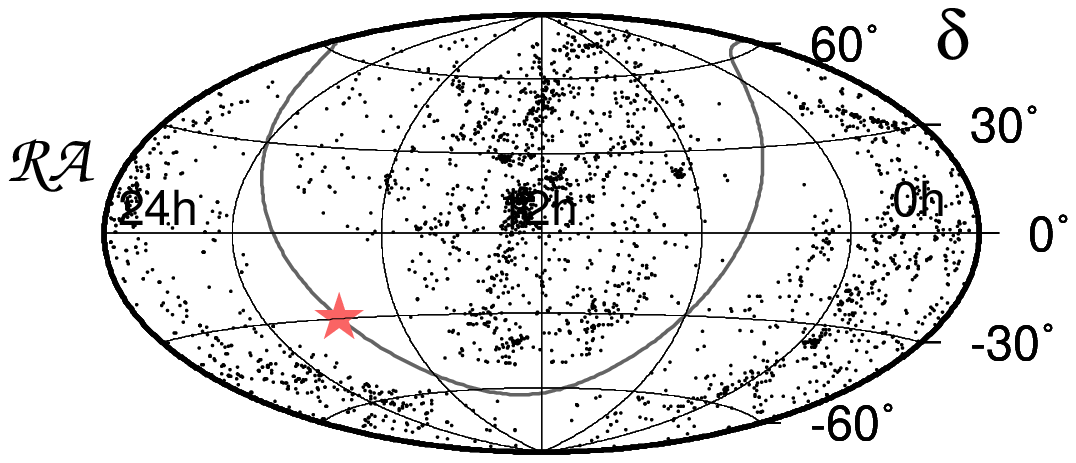}\\
(a) & (b) \\
\includegraphics[width=7.5cm,height=4cm]{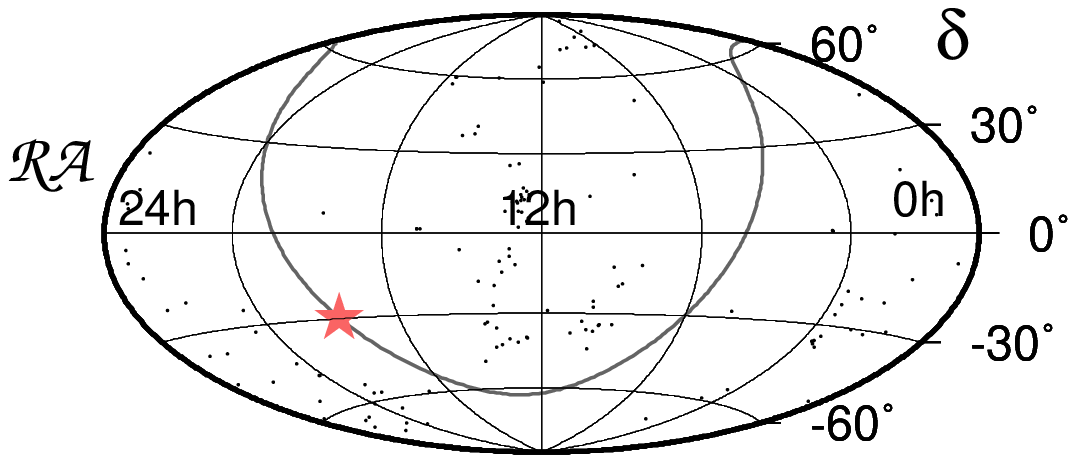}&
\includegraphics[width=7.5cm,height=4cm]{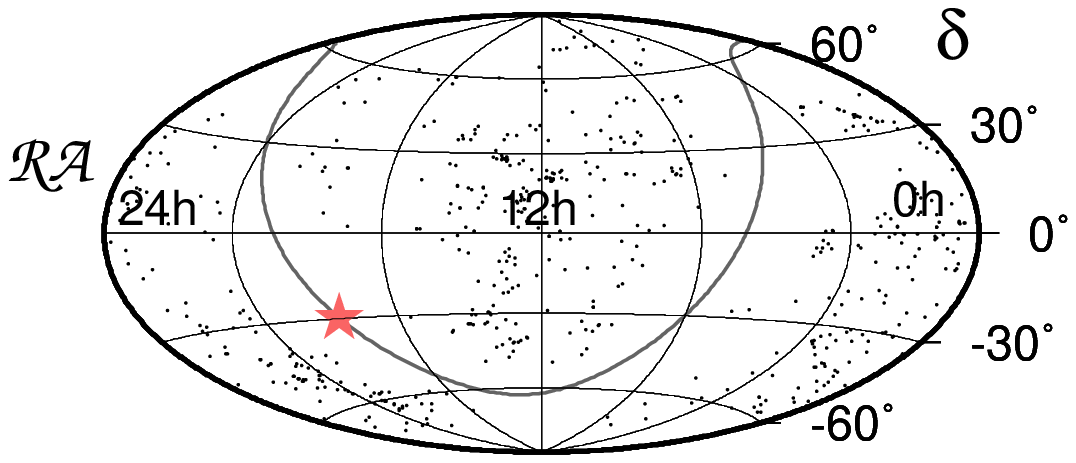}\\
(c) & (d)
\end{tabular}
\end{center}
\caption{Hammer-Aitoff projections in equatorial coordinates-s of the
four samples of galaxies, namely (a) Sample I, (b) Sample II, (c)
Sample III and (d) Sample IV. These plots record the positions of the
galaxies, not the flux of incoming UHECRs. The Galactic Center is
marked with a star.}
\label{fig:plotsofgals}
\end{figure}

\section{Nearby Extragalactic Sources}

\subsection{Models}

There are two serious impediments to the proposal that extragalactic
sources supply the UHECRs. First, most of the UHECRs do not point back
towards any obvious astrophysical sources, such as the active or
interacting galaxies within $\sim 50$ Mpc. For example, the possible
source of the highest energy Fly's Eye event was examined in
[\cite{es95}] and the best candidate amongst nearby sources was the
starburst galaxy M82, some $37^\circ$ away.  Second, the possibilities
for acceleration of cosmic rays to such extreme energies in any
extragalactic object seem to be intrinsically limited
[\cite{egsource}]. Despite the obstacles, the idea that UHECRs
originate in extragalactic sources is still worth examining. At
energies exceeding $E \sim 4 \times 10^{19}$ eV, the UHECRs in the
Haverah Park dataset exhibit a correlation with the direction of the
supergalactic plane [\cite{stanev}], though this is not seen in the
AGASA experiment [\cite{hayashida}]. However, two of the AGASA
clusters (C1 and C2) do lie close to the supergalactic plane and do
have tentative identifications with nearby galaxies.

Let us suppose that supermassive black holes in galactic nuclei
produce UHECRs. There is increasing evidence that many, perhaps all,
galaxies contain central black holes, even if their nuclei display
only low levels of activity today. Very convincing candidates are
provided by the nearby galaxy NGC 4258, where maser emission has
traced out the gas cloud motions around a central black hole of $\sim
3.6 \times 10^7 \msun$, as well as the Galaxy, where the proper
motions of stars very close to the center strongly suggest a black
hole of mass $\sim 2 \times 10^6 \msun$ [\cite{rees2}].  Stellar
kinematics can now be obtained at high spatial resolution (down to
$1''$) with the new generation of spectrographs, enabling the nuclear
regions of nearby galaxies to be probed. This has strengthened the
evidence for supermassive black holes in a number of nearby galaxies,
such as M31, M32 and M87 [\cite{richstone}]. The case of M32 is
particularly interesting. Although it is only a low luminosity galaxy,
there is strong evidence for a supermassive black hole of $\sim 2
\times 10^6\msun$ [\cite{m32}]. Only its relative closeness enables
the black hole to be identified.

The {\it Third Reference Catalogue of Bright Galaxies} (RC3) is
reasonably complete for nearby galaxies [\cite{devauc}].  It aims to
include all galaxies with total B-band magnitudes $B_{\rm T}$ brighter
than $\sim 15.5$ and with a redshift less than $15000$ \kms.  We use
it to construct the following datasets.  {\em Samples I} and {\em II}
contain all galaxies in the RC3 intrinsically brighter than M32 (NGC
221) and closer than 50 Mpc and 100 Mpc respectively.  M32 is the
smallest galaxy for which there is good evidence for a central black
hole, so it is a reasonable supposition that all galaxies larger than
M32 contain black holes, even if quiescent.  {\em Samples III} and
{\em IV} contains all galaxies in the RC3 intrinsically brighter than
Centaurus~A (NGC 5128) and closer than 50 Mpc and 100 Mpc
respectively.  Centaurus~A is the nearest active galaxy, so these
Samples contain only the brightest and largest of the nearby galaxies.

The motivation behind these choices is to answer the following
questions. What is the expected anisotropy signal if only big, bright
galaxies supply the UHECRs, or if all nearby galaxies contribute
equally?  Is it only galaxies within 50 Mpc that dominate the signal?

\begin{table}
\begin{center}
{\bf (a) AGASA}
\end{center}
\begin{center}
\begin{tabular}{|c|c|c|c|c|c|} \hline
Model & Amplitude & Phase & $P (\sigma_\S, \sigma_\vartheta)$
& $P(\sigma_\S)$ & $P(\sigma_\vartheta)$ \\ 
\null & $\S$ & $\vartheta$ & $(95 \%)$ & $(95 \%)$ & $(95 \%)$  \\ \hline 
Sample I  &0.309 &$111.1^\circ$ &0.343 &0.755 &0.523\\ \hline
Sample II &0.307 &$110.9^\circ$ &0.346 &0.759 &0.524\\ \hline 
Sample III&1.902 &$10.7^\circ$  &$<10^{-3}$&$<10^{-3}$ &0.910 \\ \hline
Sample IV &1.862 &$10.9^\circ$  &$<10^{-3}$&$<10^{-3}$&0.910 \\ \hline
\end{tabular}
\end{center}
\begin{center}
{\bf (b) Haverah Park}
\end{center}
\begin{center}
\begin{tabular}{|c|c|c|c|c|c|} \hline
Model & Amplitude & Phase & $P (\sigma_\S,\sigma_\vartheta)$
& $P(\sigma_\S)$ & $P(\sigma_\vartheta)$ \\ 
\null & $\S$ & $\vartheta$ & $(95 \%)$ & $(95 \%)$ & $(95 \%)$ \\ 
\hline 
Sample I  &0.353 &$110.8^\circ$ &0.783 &0.952 &0.828\\ \hline
Sample II &0.352 &$110.7^\circ$ &0.782 &0.952 &0.828\\ \hline 
Sample III&1.922 &$10.8^\circ$ &$<10^{-3}$ &$<10^{-3}$&0.072\\ \hline
Sample IV &1.887 &$10.9^\circ$ &$<10^{-3}$&$<10^{-3}$&0.072\\ \hline
\end{tabular}
\end{center}
\caption{For each of the four samples, the theoretical values of the
amplitude and the phase are calculated using the response functions
for (a) AGASA and (b) Haverah Park. We also give the probabilities
that the experimental amplitude and phase come from within the $95 \%$
confidence limits of the model.}
\label{table:tab6}
\end{table}
\begin{table}
\begin{center}
{\bf (a) Auger I}
\end{center}
\begin{center}
\begin{tabular}{|c|c|c|c|} \hline
Model & $P(\sigma_\S, \sigma_\vartheta)$ & $P(\sigma_\S)$ 
& $P(\sigma_\vartheta)$\\ 
\null & ($95 \%$) & ($95 \%$) & ($95 \%$) \\ \hline 
Sample I  &$<10^{-3}$ & $0.801$ &$<10^{-3}$\\ \hline
Sample II  &$<10^{-3}$ & $0.811$ &$<10^{-3}$\\ \hline
\end{tabular}
\end{center}
\begin{center}
{\bf (b) Auger II}
\end{center}
\begin{center}
\begin{tabular}{|c|c|c|c|c|c|c|} \hline
Model 
& $P(\sigma_\S, \sigma_\vartheta)$ & $P(\sigma_\S)$ & $P(\sigma_\vartheta)$\\ 
\null&($95 \%$)&($95 \%$) & ($95 \%$) \\ \hline 
Sample I  &$<10^{-3}$ & $0.001$ &$<10^{-3}$\\ \hline
Sample II  &$<10^{-3}$ & $0.001$ &$<10^{-3}$\\ \hline
\end{tabular}
\end{center}
\begin{center}
{\bf (c) Predictions for South Auger}
\end{center}
\begin{center}
\begin{tabular}{|c|c|c||c|c|}\hline
Model&Amplitude &Phase &Amplitude &Phase \\ 
\null& $(h_1)$ & $(h_1)$ & $(h_2)$& $(h_2)$\\ \hline
Sample I  &0.471 &$49.4^\circ$ &1.187 &$22.4^\circ$\\ \hline
Sample II &0.469 &$48.7^\circ$ &1.178 &$22.3^\circ$\\ \hline
Sample III&0.568 &$123.2^\circ$ &1.395 &$85.5^\circ$\\ \hline
Sample IV &0.411 &$116.0^\circ$ &1.268 &$83.9^\circ$\\ \hline
\end{tabular}
\end{center}
\caption{For each of the Samples, this shows the impact of the South
Auger Observatory. Table (a) assumes the \ch{amplitude and phase} of
the AGASA experiment, table (b) the \ch{amplitude and phase} from
Haverah Park; in both cases, 1000 events are analysed. In Tables (a)
and (b), lines are omitted if all the probabilities vanish.  Table (c)
presents the predictions for our four Samples using the two possible
response functions $h_1(\delta)$ and $h_2(\delta)$.}
\label{table:tab7}
\end{table}
\begin{figure}
\begin{center}
\begin{tabular}{c@{\hspace{2cm}}c}
  \psfrag{r}{$\S$}
  \psfrag{p}{$\vartheta$}
  \includegraphics[width=7cm,height=7cm]{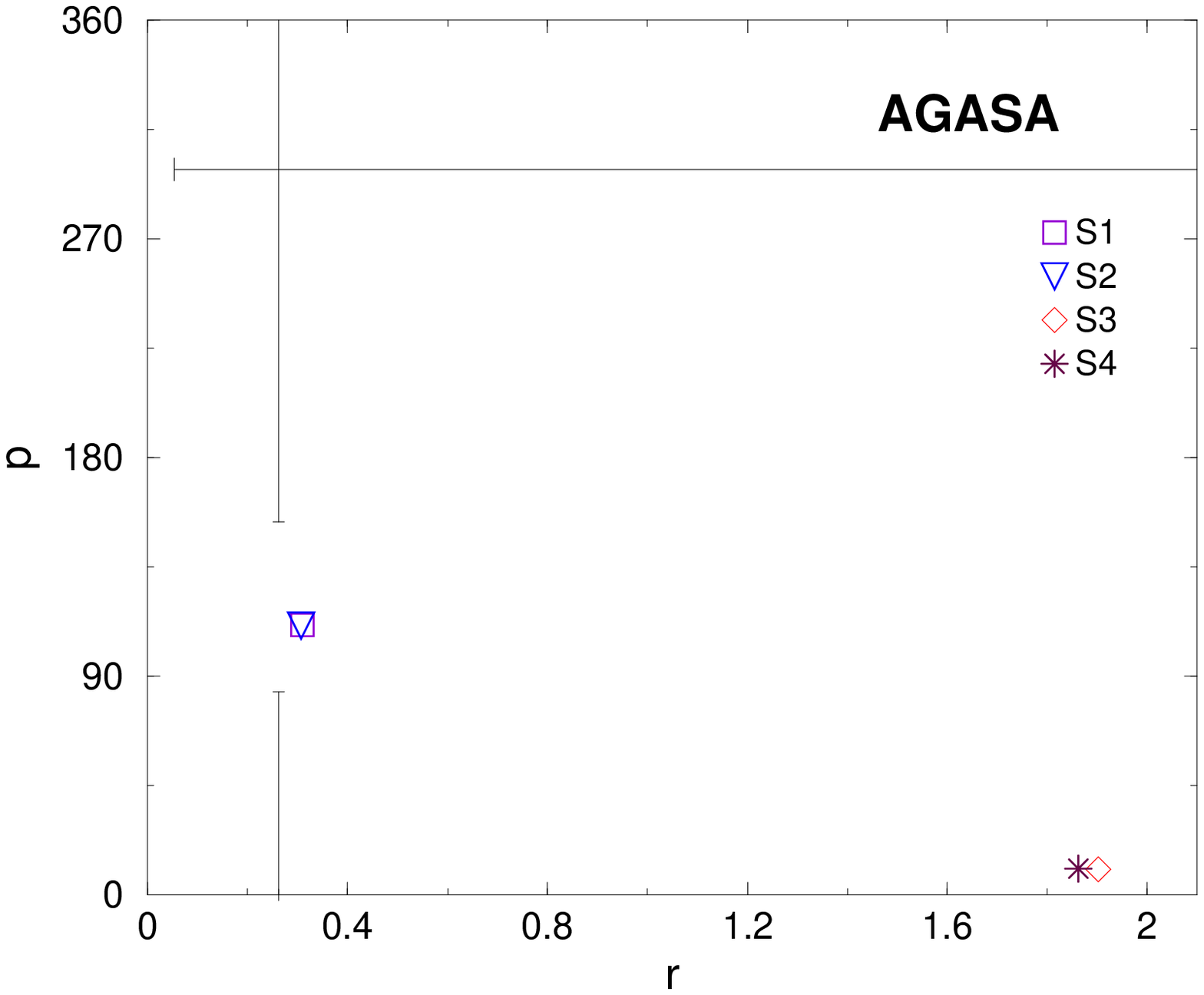}&
  \psfrag{r}{$\S$}
  \psfrag{p}{$\vartheta$}
  \includegraphics[width=7cm,height=7cm]{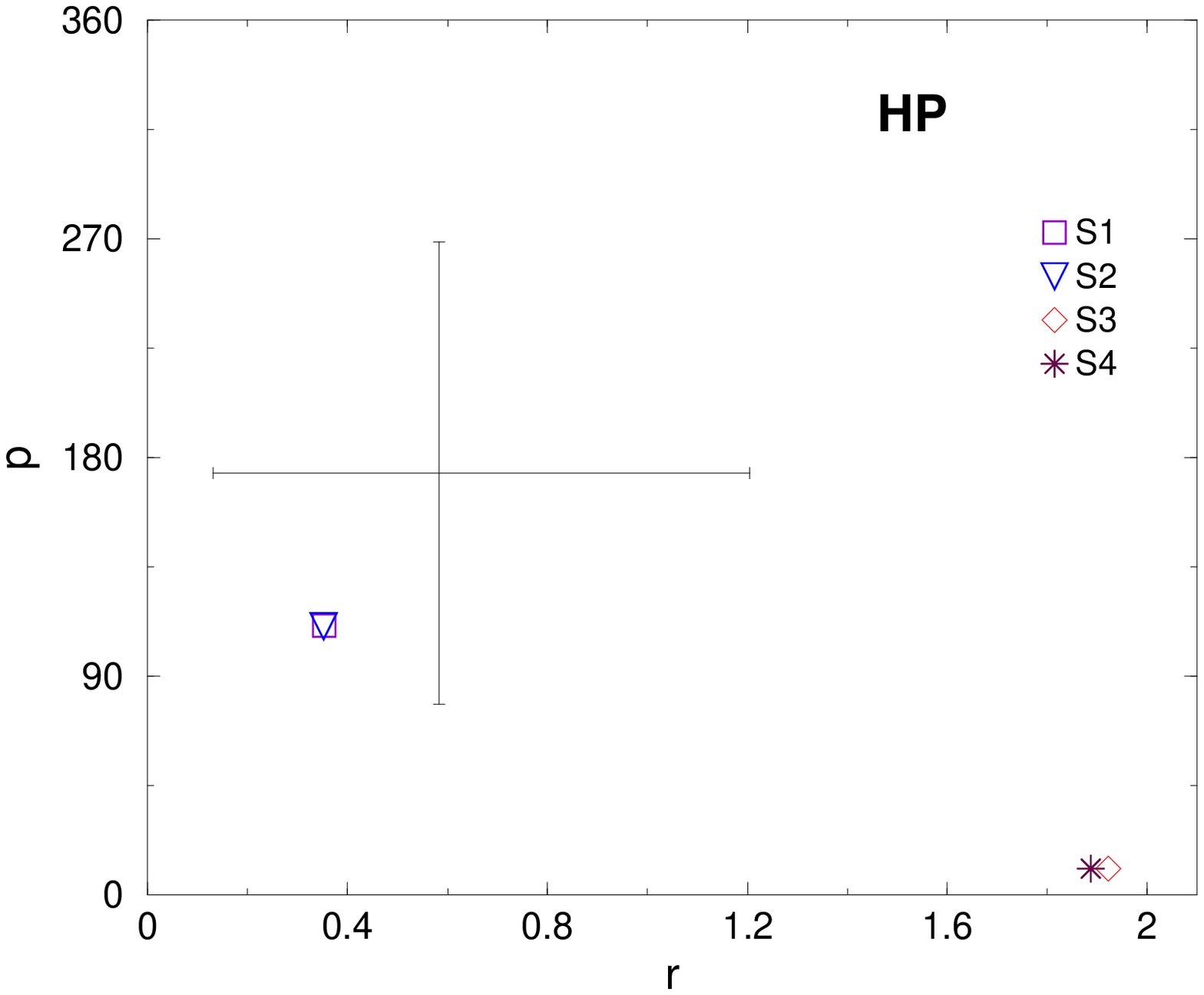}\\
(a) & (b) \\
&\\
&\\
  \psfrag{r}{$\S$}
  \psfrag{p}{$\vartheta$}
  \includegraphics[width=7cm,height=7cm]{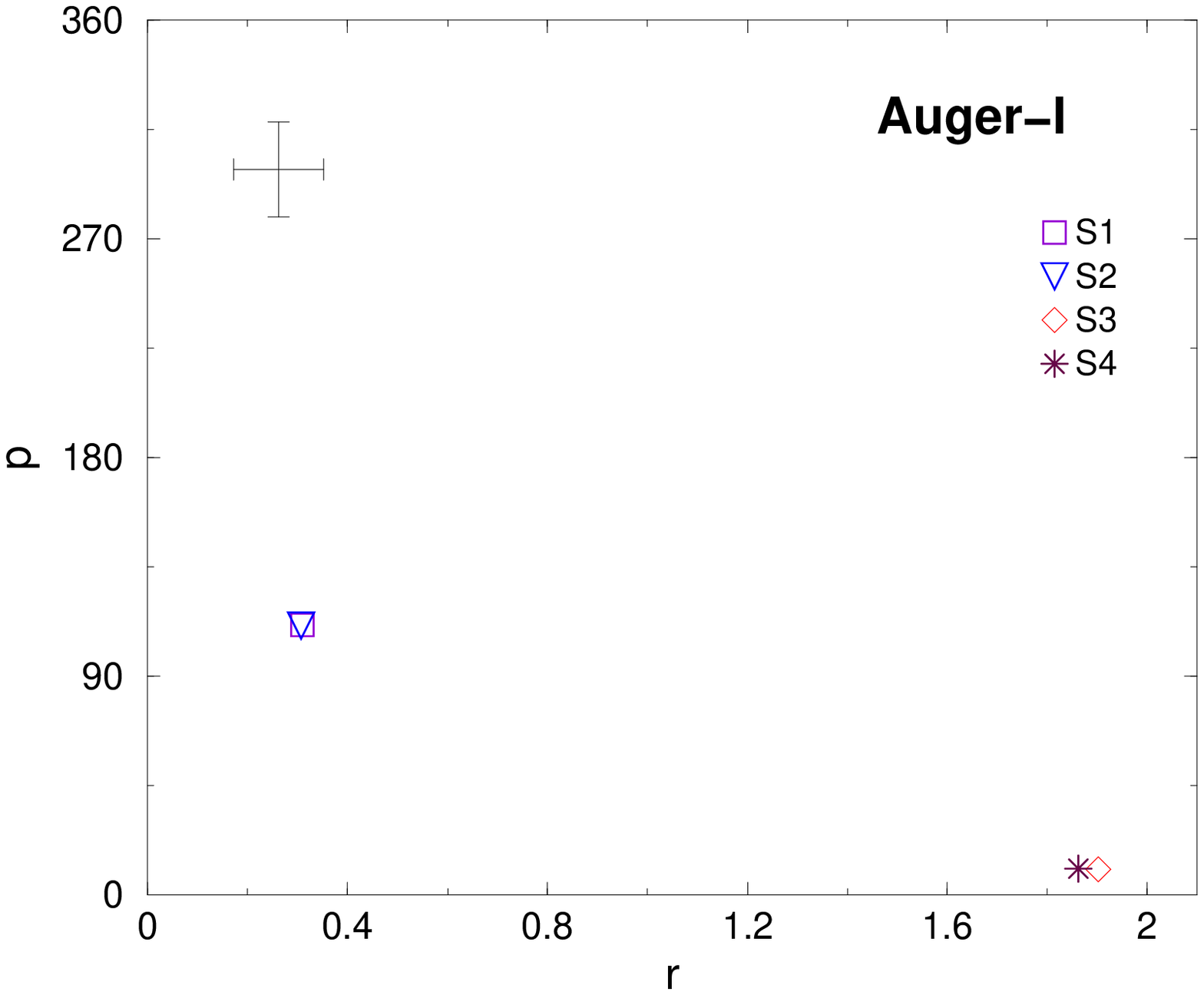}&
  \psfrag{r}{$\S$}
  \psfrag{p}{$\vartheta$}
  \includegraphics[width=7cm,height=7cm]{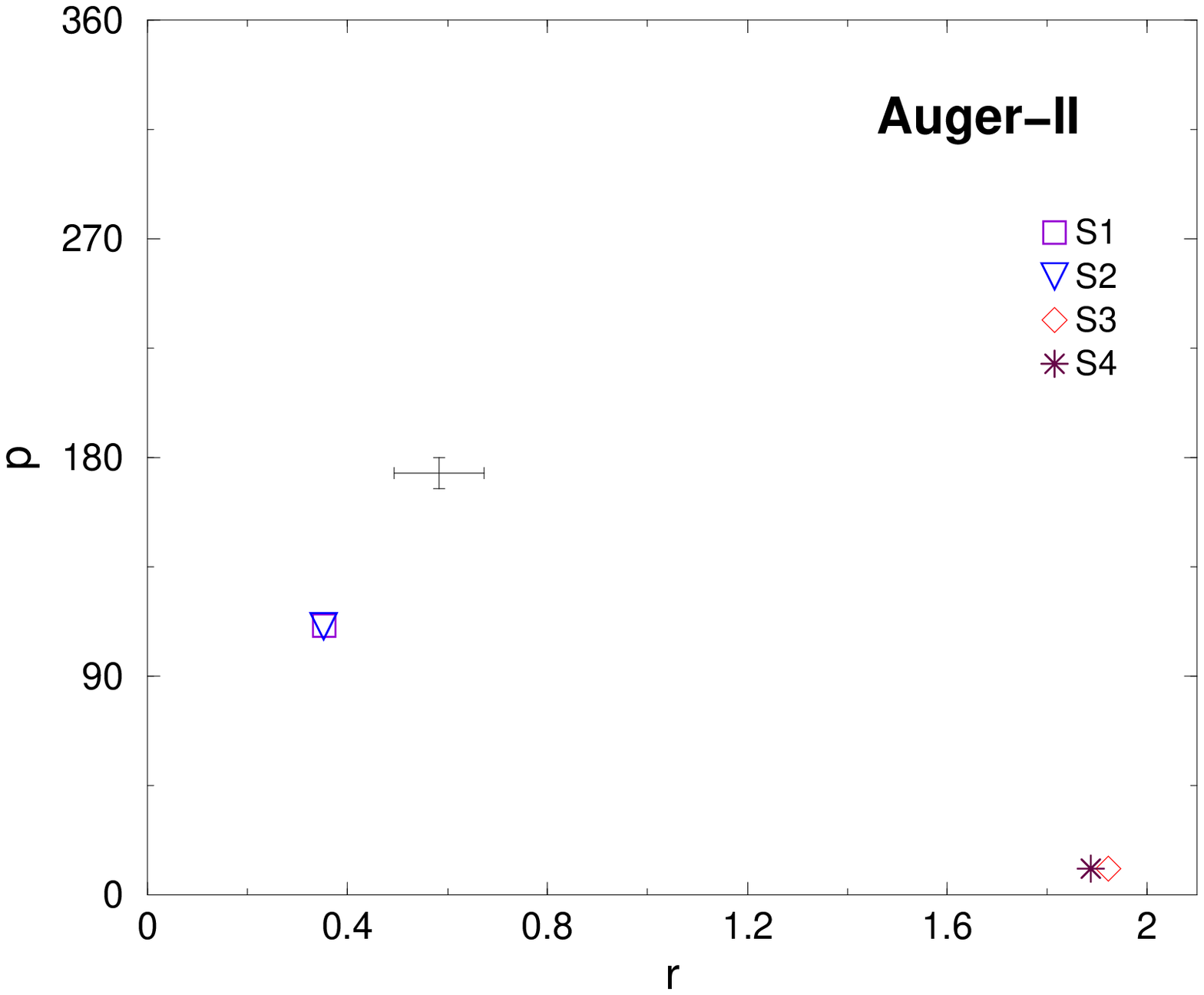}\\
(c) & (d) 
\end{tabular}
\end{center}
\caption{Plots of the amplitude against the phase for the four Samples
of galaxies. Also shown are the $95 \%$ confidence limits for (a) the
AGASA and (b) the Haverah Park experiments. The lower panels show the
expected impact of South Auger. Marked are the $95 \%$ confidence limits
assuming the amplitude and phase of (c) the AGASA and (d) the Haverah
Park experiments, but with 1000 events.}
\label{fig:harmonicgalaxies}
\end{figure}

\subsection{Results}

To take into account the declinations of the galaxies, we use the
discrete analogue-s of equations~(\ref{eq:han0}) and ~(\ref{eq:han}):
\begin{equation}
N = \sum_i h(\delta_i) {1 \over r_i^2 },
\label{eq:ffone}
\end{equation}
\begin{equation}
a = \frac{2}{N} \sum_i h(\delta_i) {\cos \alpha_i \over r_i^2},\qquad
b = \frac{2}{N} \sum_i h(\delta_i) {\sin \alpha_i \over r_i^2},
\label{eq:fftwo}
\end{equation}
where the index $i$ runs along all the galaxies in the Sample and
$h(\delta)$ is the response function of each experiment.
 
Figure~\ref{fig:plotsofgals} shows the distribution of galaxies in the
four Samples. The supergalactic plane is clearly visible, as is the
Zone of Avoidance within $\sim 15^\circ$ of the Galactic plane.  Some
of the prominent local features are also readily apparent.  The Virgo
cluster is centered at $\alpha \approx 12^{\rm h} 30^{\rm m}, \delta
\approx 10^\circ$, while the Fornax cluster is at $\alpha \approx
3^{\rm h}, \delta \approx -40^\circ$. The densest part of the Local
Supercluster and the Hydra-Centaurus supercluster are prominent at
$\alpha \approx 12^{\rm h} - 1^{\rm h}, \delta \approx -40^\circ$,
while the Perseus-Pisces cluster stretches from $\alpha \approx 1^{\rm
h} - 3^{\rm h}, \delta \approx 40^\circ$. Of course, any optically
selected survey is strongly affected by extinction at low Galactic
latitudes. We could avoid this bias by using a different catalogue
such as the {\it IRAS} galaxies. This though merely substitutes one
bias for another, as early-type galaxies are under-represented in the
{\it IRAS} survey, whereas late-type dusty disk galaxies are
over-represented.  How serious is the bias caused by low latitude
extinction on our calculations?  Let us note that the Dwingeloo
Obscured Galaxy Survey conducted a shallow search in the entire
Northern Zone of Avoidance for nearby and/or massive galaxies and this
yielded just 5 candidates [\cite{dwingeloo}].  They estimated that
there are $\sim 100$ galaxies missing within $\sim 50$ Mpc. The
distance factors in eqs.~(\ref{eq:ffone},\ref{eq:fftwo}) mean that the
nearest galaxies have the largest weight. Hence, this suggests that
low latitude obscuration will not have a deleterious effect on our
results. Only if there are large numbers of missing galaxies within 10
Mpc is our calculation likely to be in error, and this circumstance
does not seem to be the case, as judged by the available evidence.

Table~\ref{table:tab6} shows the signal expected for AGASA and Haverah
Park. For all four Samples, the size of the amplitude is significant,
so that detection of the anisotropy is not enough to implicate
decaying dark matter haloes.  Samples I and II, which contain almost
all the nearby galaxies, do yield an anisotropic signal of the same
order of magnitude as seen by AGASA and Haverah Park.  In fact, the
direction of the signal is well within the $95 \%$ confidence limits
for the Haverah Park dataset.  The probabilities that the AGASA and
Haverah Park datasets are drawn from within the $95 \%$ confidence
limits of the model are moderate ($0.264$) and high ($0.858$)
respectively.  The phase of the anisotropy is robust against changes
in the cut-off radius, suggesting that it is the very closest galaxies
(within $\sim 20$ Mpc) that are contributing most of the signal. The
direction of the signal shows that it is is dominated by nearby
structures in the supergalactic plane, especially the Virgo
cluster. Of all our models, Samples III and IV, which contain the
nearby bright galaxies within 50 and 100 Mpc respectively, give the
largest amplitude for the anisotropy signal.  They are clearly
disfavored by the existing AGASA and Haverah Park datasets.
Table~\ref{table:tab6} shows there is vanishingly small probability
that either dataset is drawn from within the $95 \%$ confidence limits
of the model.  The observed distribution of UHECRs seems to be more
isotropic than expected if they emanate just from the nearby bright
galaxies.

Table~\ref{table:tab7} shows the potential effects of the Auger
Observatory. Again, Auger I has the same amplitude and phase as
recorded by AGASA, while Auger II is the same as that \ch{recorded} by
Haverah Park, with the number of events increased to 1000. Of course,
if UHECRs originate in nearby galaxies, then the Northern hemisphere
signal will be different from the Southern hemisphere. We are simply
resizing the experimental box to match Auger's sensitivity as an
illustration of its ability to discriminate.  Tables~\ref{table:tab7}
(a) and (b) are composed mainly of zeros. The large number of events
now enables all four Samples to be ruled out, as the signals do not
coincide with those measured by AGASA and Haverah
Park. Figure~\ref{fig:harmonicgalaxies} shows these results
pictorially, together with the $95 \%$ confidence limits.

The signal seen by South Auger is recorded in Table~\ref{table:tab7}
(c). The amplitude is greater than that seen by the Northern
Hemisphere detectors, and the phase points in a different
direction. For Samples I and II, we obtain results that depend on the
assumed detector response so it is difficult to make definite
predictions. The number of events for a $5\sigma$ detection in the
amplitude is $\sim 120$ (for the $h_1$ response) and $\sim 40$ (for
the $h_2$ response).  For Samples III and IV, we obtain large
amplitudes irrespective of the response function. Only $\sim 30$
events are needed for a $5\sigma$ detection of the amplitude.  The
direction of the phase is largely controlled by the Fornax cluster,
which is the nearest rich cluster to us. Although it is less massive
than the Virgo cluster, its effect is dominant in the Southern
sky. Suppose that we wished to ensure that the $3\sigma$ (or
$5\sigma$) error bars on the measured phase at South Auger are $\pm
20^\circ$. This is sufficient to implicate an extragalactic origin of
the UHECRs and requires $\sim 350$ (or $\sim 1000$) events.

\section{Conclusions}

The origin of the ultra-high energy cosmic rays (UHECRs) is unknown,
but the distribution of arrival directions provides important
clues. We have examined two possible hypotheses for the origin in
detail. As the available dataset of UHECRs is too small to yield
definitive evidence for anisotropy, we have concentrated on the
prospects for the Auger Observatory, especially its southern station
at Malarg{\"u}e, Argentina.

First, UHECRs may be produced by the decay of dark matter in the halo
of the Galaxy. The offset of the Sun from the Galactic Centre causes
an anisotropy signal. The magnitude of this anisotropy is largest for
dark halo models which are cusped (such as the Navarro-Frenk-White or
NFW profile). However, there is overwhelming astronomical evidence
that the Galactic halo does not have a cusped density profile. It is
more likely that the halo has a core radius $\sim 10$ kpc, so that
dark matter is dynamically unimportant in the central parts but
dominates more and more beyond the Solar circle.  The halo may be
isothermal or triaxial or possibly even tilted with respect to the
Galactic plane. If such models describe the distribution of decaying
dark matter, then the amplitude of the anisotropy $\S$ is $\lesssim
0.4$ for the AGASA and Haverah Park detectors, whilst the phase of the
anisotropy $\vartheta$ points towards the direction of the Galactic
Centre. The shape of the halo controls the phase, whereas the scale of
the halo controls the amplitude. Spherical halo models yield
anisotropies whose phase coincides with the direction of the Galactic
Centre, whereas triaxial models give angular deviations of $\sim
25^\circ$ depending on the halo shape and detector location. The
amplitude of the anisotropy is controlled by the halo core radius
$\rc$. This parameter is not easy to determine by astrophysical
means. So, should this mechanism of UHECR production be proved correct
(e.g., through confirmation of the expected energy spectrum
[\cite{bs98}]), then it will provide a direct way of measuring the
core radius.

If UHECRs originate in dark haloes, then it has been suggested that
the halo of M31 would be visible as a hotspot~[\cite{bsw99}].
However, the size of this effect seems to be modest. Partly this is
because recent analyses have revised the overall mass of the M31 halo
downwards, making it roughly comparable to the mass of the Galactic
halo~[\cite{ewggv}]. Furthermore, UHECRs emanating from M31 are
deflected by a few degrees and so the effect must be sought in a field
size of $\sim 10^\circ \times 10^\circ$, which dilutes the
signal~[\cite{mtw99}]. Our calculations suggest that only a mild
enhancement is expected ($\lesssim 60 \%$) and its identification
requires considerably more events than in the AGASA and Haverah Park
datasets, in agreement with~[\cite{mtw99}] but not
with~[\cite{bsw99}]. Note that the South Auger detector will not help
much in this regard, as M31 is not visible from the latitude of
Malarg{\"u}e.

A robust prediction of the decaying dark matter hypothesis is that the
amplitude of the anisotropy at the South Auger station will be larger
than at the Northern Hemisphere sites, simply because the Galactic
Centre lies at southern declinations. Isothermal haloes, whether
spherical, triaxial or tilted, yield anisotropies of amplitude $\sim
0.5$ for a halo core radius $\sim 10$ kpc.  Typically, the detection
of between 150 and 500 events at South Auger will be required for a $5
\sigma$ identification of the anisotropy signal. However, if the core
radius is smaller ($\lesssim 1$ kpc), then the amplitude of the
anisotropy is $\gtrsim 1.0$ and perhaps as few as 40 events would
suffice. The phase will give information on triaxiality or tiltedness,
and this may be obtained with $\sim 2000$ events for isothermal-like
models. More importantly, $\sim 500$ events are sufficient to confirm
that the phase points in the rough direction of the Galactic Center,
which would implicate a Galactic origin of the UHECRs.

A second hypothesis is that the UHECRs may originate in the nuclei of
nearby galaxies, perhaps produced by supermassive black holes.  If
only the nearby bright galaxies are the sources, then the amplitude of
the anisotropy is much greater than observed in the AGASA and Haverah
Park datasets (typically $\S \sim 1.9$). It therefore seems that this
hypothesis can already be ruled out, as the probabilities that the
AGASA and Haverah Park datasets are consistent with such signals are
very small. However, some caution is needed as the number of UHECRs is
still small and the effects of low latitude obscuration have been
neglected in the analysis. Furthermore, the expected anisotropy may be
diluted by up to a factor of $\sim 2$ by a possible isotropic
background.

If extragalactic sources are to provide the UHECRs, then the
population must be larger than just the nearby bright galaxies.  The
smallest galaxy presently suspected of possessing a supermassive black
hole is M32. Samples of nearby galaxies brighter than M32 yield an
anisotropy signal $\S \sim 0.5$ in amplitude and $\vartheta \sim
120^\circ$ in phase.  This is in good agreement with the signal seen
by Haverah Park, and in rough agreement with that seen by AGASA. In
the decaying dark matter hypothesis, the phase always points in the
approximate direction of the Galactic Centre. If, however, all nearby
galaxies provide the UHECRs, then the phase at the AGASA and Haverah
Park detectors points towards $\alpha \approx 9^{\rm h}$ and is
controlled by prominent mass concentrations in the supergalactic
plane, such as the Virgo cluster ($\alpha \approx 12^{\rm h} 30^{\rm
m}, \delta \approx 10^\circ$).

The measurement of significant anisotropy with the South Auger station
does not by itself validate the decaying dark matter hypothesis, as
this phenomenon occurs in our extragalactic samples as well.  When all
galaxies intrinsically brighter than M32 are included, the amplitude
of the anisotropy is $\S \sim 0.6$ irrespective of whether the cut-off
is 50 or 100 Mpc.  The number of events for a $5\sigma$ detection in
the amplitude is $\lesssim 120$.  The phase $\vartheta$ does not point
toward the Galactic Centre, but is controlled by the nearby Fornax
cluster ($\alpha \approx 3^{\rm h}, \delta \approx -40^\circ$).
Suppose that we wished to ensure that the $3\sigma$ error bar on the
measured phase at South Auger is $\pm 20^\circ$, which would be strong
evidence for an extragalactic origin. This requires $\sim 350$ events.

In conclusion, it seems that one of the most important contributions
that the South Auger experiment can make is to identify the phase
direction in the Southern hemisphere.  If UHECRs have a Galactic
origin, then the phase will point towards the Galactic Centre. If
UHECRs have an extragalactic origin, then it seems they must be
produced by almost all nearby galaxies (else the signal seen by AGASA
and Haverah Park would be much larger). Then, the phase of the
anisotropy recorded by South Auger will lie roughly in the direction
of the Fornax cluster. Provided the effects of the intergalactic
magnetic field can be neglected, this gives an unambiguous
discriminant between the two theories and requires only $\sim 350-500$
events. This should be obtained within the first three years of
operation of South Auger.

\bigskip
\noindent
{\bf Acknowledgments}: We thank Motohiko Nagano for providing the
experimental datasets, as well as Paul Sommers and Alan Watson for a
number of helpful discussions and clarifications. The anonymous
referees provided a careful reading of the manuscript and made a
number of helpful suggestions.  NWE is supported by the Royal
Society.  FF is partially supported by the CICYT Research Project
AEN99-0766 and the CIRIT. He thanks the sub-Department of Theoretical
Physics, Oxford for hospitality extended to him during working visits.


\end{document}